\documentclass[]{myaa}
\usepackage{graphicx}
\usepackage[varg]{txfonts}
\usepackage{natbib}
\bibpunct{(}{)}{;}{a}{}{,} % to follow the A&A style
\usepackage[figuresright]{rotating}
\usepackage[a4paper,breaklinks,dvipdfm]{hyperref}

\def\teff{$T\rm_{eff}$}
\def\kms{$\mathrm{km\, s^{-1}}$}

\newcommand{\logg}{\ensuremath{\log {\rm g}}}

\newcommand{\beq}{\begin{equation}}
\newcommand{\eeq}{\end{equation}}

\begin{document}
\idline{100}

\title{TOPoS: II. On the bimodality of carbon abundance in CEMP stars \thanks{Based
on observations obtained at ESO Paranal Observatory,
programme 091.D-0288, 091.D-0305, 189.D-0165.}
}
\subtitle{Implications on the early chemical evolution of galaxies}

\author{
P.~Bonifacio   \inst{1} \and
E.~Caffau\thanks{MERAC Fellow}\inst{1,2} \and
M.~Spite       \inst{2} \and
M.~Limongi   \inst{3,4}\and
A.~Chieffi   \inst{5}\and
R.S.~Klessen     \inst{6,7,8} \and
P.~Fran\c{c}ois \inst{1,9} \and
P.~Molaro      \inst{10}\and
H.-G.~Ludwig  \inst{2,1} \and
S.~Zaggia     \inst{11} \and
F.~Spite    \inst{1} \and
B.~Plez     \inst{12}
R.~Cayrel     \inst{1} \and
N.~Christlieb \inst{2} \and
P.C.~Clark      \inst{13,6} \and
S.C.O.~Glover      \inst{6} \and
F.~Hammer      \inst{1} \and
A.~Koch       \inst{2} \and
L.~Monaco      \inst{14,15} \and    
L.~Sbordone    \inst{16,17} \and
M.~Steffen     \inst{18,1} 
}

\institute{ 
GEPI, Observatoire de Paris, PSL Resarch University, CNRS, Univ Paris Diderot, 
Sorbonne Paris Cit\'e, Place Jules Janssen, 92195 Meudon, France
\and
Zentrum f\"ur Astronomie der Universit\"at Heidelberg, Landessternwarte, 
K\"onigstuhl 12, 69117 Heidelberg, Germany
\and
Istituto Nazionale di Astrofisica,
Osservatorio Astronomico di Roma, via Frascati 33, I-00040 Roma, Italy
\and
Kavli Institute for the Physics and Mathematics of the Universe, Todai
Institutes for Advanced Study, The University of Tokyo, Kashiwa 277-8583,
Japan
\and
Istituto Nazionale di Astrofisica,
Istituto di Astrofisica Spaziale e Fisica Cosmica, 
Via Fosso del Cavaliere 100, I-00133 Roma, Italy
\and
Zentrum f\"ur Astronomie der Universit\"at Heidelberg,
Institut f\"ur Theoretische Astrophysik, Albert-Ueberle-Stra{\ss}e 2, 69120 Heidelberg, Germany
\and
Kavli Institute for Particle Astrophysics and Cosmology, Stanford University, SLAC National Accelerator Laboratory, Menlo Park, CA 94025, USA
\and
Department of Astronomy and Astrophysics, University of California, 1156 High Street, Santa Cruz, CA 95064, USA
\and
UPJV, Universit\'e de Picardie Jules Verne, 33 Rue St Leu, F-80080 Amiens
\and
Istituto Nazionale di Astrofisica,
Osservatorio Astronomico di Trieste,  Via Tiepolo 11,
I-34143 Trieste, Italy
\and
Istituto Nazionale di Astrofisica,
Osservatorio Astronomico di Padova Vicolo dell'Osservatorio 5, 35122 Padova, Italy
\and
Laboratoire Univers et Particules de Montpellier, LUPM, Universit\'e Montpellier, 
CNRS, 34095 Montpellier cedex 5, France
\and
School of Physics and Astronomy, The Parade, Cardiff University, Cardiff, CF24 3AA
\and
European Southern Observatory, Casilla 19001, Santiago, Chile
\and
Departamento de Ciencias F\'isicas, Universidad Andr\'es Bello, Rep\'ublica 220, 837-0134 Santiago, Chile
\and
Millennium Institute of Astrophysics,
Vicu{\~n}a MacKenna 4860, Macul, Santiago, Chile
\and
Pontificia Universidad Cat{\'o}lica de Chile
Vicu{\~n}a MacKenna 4860, Macul, Santiago, Chile
\and
Leibniz-Institut f\"ur Astrophysik Potsdam (AIP), An der Sternwarte 16, 14482 Potsdam, Germany
}
\authorrunning{Bonifacio et al.}
\titlerunning{TOPoS: II.  C-enhanced  stars}
%\offprints{P.~Bonifacio}
%\date{Received ...; Accepted ...}

\abstract%
%\context
{
In the course of the TOPoS (Turn Off Primordial Stars) survey,  aimed at discovering the lowest
metallicity stars, we have found several carbon-enhanced metal-poor (CEMP) stars.
These stars are very common among the stars of extremely low metallicity
and provide important clues to the star formation processes.
We here present our analysis of six CEMP stars.}
%\aims
{
We want to provide the most complete chemical inventory for 
these six 
stars in order to constrain the nucleosynthesis processes
responsible for the abundance patterns.
}
%\method
{
We analyse both X-Shooter and UVES spectra acquired at  the VLT.
We used a traditional abundance analysis
based on 
OSMARCS 1D Local Thermodynamic Equilibrium (LTE) model atmospheres and the {\tt turbospectrum}
line formation code. 
}
%\results
{
Calcium and carbon are  the only elements that
can be measured in all six stars. The
range is $-5.0\le$[Ca/H]$< -2.1$ and
$7.12\le$A(C)$\le 8.65$. 
For star SDSS\,J1742+2531 
we were able to detect three \ion{Fe}{i} lines
from which we deduced $[{\rm Fe/H}]=-4.80$,
from four \ion{Ca}{ii} lines we derived
[Ca/H]=--4.56, and from synthesis
of the G-band we derived A(C)=7.26.
For SDSS\,J1035+0641 we were  not able to detect any
iron lines, yet we could place a robust (3$\sigma$) upper limit
of $[{\rm Fe/H}]< -5.0$ and measure the Ca abundance,
with [Ca/H]=--5.0, and carbon, A(C)=6.90,  suggesting that this star
could be even more metal-poor than  SDSS\,J1742+2531.
This makes  
these two stars the seventh and eighth stars known so far
with $[{\rm Fe/H}]<-4.5$, usually termed ultra-iron-poor (UIP) stars. No lithium is detected
in the spectrum of SDSS\,J1742+2531 or SDSS\,J1035+0641, which  
implies a robust upper limit of A(Li)$<1.8 $ for both stars.
}
%\conclusions 
{ 
Our measured carbon abundances  confirm the bimodal distribution
of carbon in CEMP stars, identifying
a {\em high-carbon band} and a {\em low-carbon band}. 
We propose an interpretation of this
bimodality according to which the stars on the {\em high-carbon band} 
are the result of mass transfer from an AGB companion, while
the stars on the {\em low-carbon band} are genuine fossil  records
of a gas cloud that has also been enriched  by a faint supernova (SN) 
providing carbon and the lighter elements.
The abundance pattern of the UIP stars shows a large star-to-star scatter in 
the [X/Ca] ratios for all elements up to aluminium (up to 1\,dex), 
but this scatter drops for heavier elements and is at most
of the order of a factor of two.
We propose that this can be explained if these stars are formed from
gas that has been chemically enriched by several SNe, that produce the
roughly constant [X/Ca] ratios for the heavier elements, and in some
cases the gas has also been polluted by the ejecta of a faint SN 
that contributes the lighter elements in variable amounts.
The absence of lithium in  three of the four known unevolved UIP stars
can be explained by a dominant role of fragmentation in the
formation of these stars. This would result either in a destruction
of lithium in the pre-main-sequence phase, through rotational
mixing or to a lack of lateaccretion from a reservoir of
fresh gas. The phenomenon should have varying degrees
of efficiency. }
\maketitle

%%%%%%%%%%%%INTRODUCTION%%%%%%%%%%%%%%%%%%%%%%%%%%%%%%%%%%%

\begin{table*}
\caption{\label{allstar}
Coordinates and 
magnitudes from SDSS}
\renewcommand{\tabcolsep}{3pt}
\tabskip=0pt
\begin{center}
\begin{tabular}{lccrrrrrrrr}
\hline\noalign{\smallskip}
\multicolumn{1}{l}{SDSS ID}& 
\multicolumn{1}{c}{RA}&
\multicolumn{1}{c}{Dec}& 
\multicolumn{1}{c}{l} &
\multicolumn{1}{c}{b} &
\multicolumn{1}{c}{$u$}& 
\multicolumn{1}{c}{$g$}& 
\multicolumn{1}{c}{$r$}&
\multicolumn{1}{c}{$i$}&
\multicolumn{1}{c}{$z$}&
\multicolumn{1}{c}{E(B--V)}\\
 &  J2000.0 & J2000.0 & \multicolumn{1}{c}{deg} & \multicolumn{1}{c}{deg} & [mag] & [mag] & [mag] & [mag] & [mag] & [mag]   \\
\noalign{\smallskip}\hline\noalign{\smallskip}
SDSS\,J0212+0137 & 02\, 12\, 38.48  & +01\, 37\, 58.08 &  160.37835821 &  $-55.22255068$ & 18.32 & 17.46 & 17.22 & 17.12 & 17.09  & $0.032$ \\
SDSS\,J0929+0238 & 09\, 29\, 12.33  & +02\, 38\, 17.00 &  230.78095136 &  35.85905110    & 19.27 & 18.36 & 17.96 & 17.78 & 17.73  & $0.061$  \\  
SDSS\,J1035+0641 & 10\, 35\, 56.11  & +06\, 41\, 43.97 &  239.11776469 &  51.90944644    & 19.53 & 18.65 & 18.37 & 18.31 & 18.26  & $0.028$  \\
SDSS\,J1137+2553 & 11\, 37\, 23.26  & +25\, 53\, 54.30 &  174.34692383 &  25.89841652    & 17.35 & 16.47 & 16.21 & 16.12 & 16.11  & $0.022$ \\  
SDSS\,J1245--0738$^a$ & 12\, 45\, 02.68  & -07\, 38\, 47.10 &  191.26116943 &  $-7.64641666$  & 17.27 & 16.31 & 16.02 & 15.91 & 15.86  & $0.026$\\
SDSS\,J1742+2531$^b$ & 17\, 42\, 59.68  & +25\, 31\, 35.90 & 49.85799222   & 25.64468718     &$20.06$ & $18.91$ & $18.67$ & $18.55$ & $18.49$ &  $0.065$ \\ 
\noalign{\smallskip}\hline\noalign{\smallskip}
\multicolumn{11}{l}{$^a$ already studied by \citet{aoki13}}\\
\multicolumn{11}{l}{$^b$ already studied by \citet{gto13}}\\
\end{tabular}
\end{center}
\end{table*}

\section{Introduction}

Carbon-enhanced stars are the most common objects in the extremely iron-poor regime.
Very few stars with iron abundance below 1/10\,000 do not show an over-abundance of carbon.
They provide important clues that help understand 
the formation of the first low-mass stars in the history of the Universe.
A carbon-enhanced metal-poor (CEMP) star has traditionally been defined
as a star having [C/Fe]$ > +1$ \citep{BC05}, although
[C/Fe]$\ge +0.7$ has 
been  used \citep[e.g.][]{Aoki07,carollo,Norris13,Lee13,Carollo14,SK15}. 
In our opinion the latter criterion should not be used, since it may
lead to ambiguous classifications. 
In fact the mean carbon-to-iron ratio
of turn-off  extremely metal-poor stars
is $\rm <[C/Fe]> = +0.45 \pm 0.1$ \citep{bonifacio09}. 
These
definitions are  always purely phenomenological, and are not
based on any theoretical consideration.
Ambiguity may arise for some very luminous giants that are expected
to have decreased their C abundance owing to mixing with material
processed through the CNO cycle \citep{SK15}.
A further shortcoming of this definition is for stars for which
Fe cannot be measured. However, in all practical cases the upper
limit on [Fe/H] allows  a lower limit on [C/Fe] to be derived which
determines whether  a star is a CEMP or not.
For the time being there are no major controversies on which stars are
to be considered CEMP, although there are some on DLAs.

The  turn-off Primordial Stars Survey  (TOPoS) \citep{topos1}
\footnote{The TOPoS survey is based mainly
on  ESO Large Programme 189.D-0165, and also on the
precursor pilot programme conducted  during the
French and Italian X-Shooter 
guaranteed time, and on some normal ESO observing 
programmes conducted with UVES such as 091.D-0288 and 091.D-0305. }
 is aimed at finding stars of extremely
low metallicity extracted from candidates selected
from spectra of the Sloan Digital Sky Survey \citep{york00,yanny09,dawson13},
data releases 7, 8, and 9 \citep{dr7,dr8,dr8e,dr9}.
Our targets are mainly 
the carbon-normal turn-off stars. However, 
in warm stars the G-band may  not
be prominent in the  low-resolution SDSS spectra,
even if the star is a CEMP. When observed at
higher resolution these stars reveal their true nature.
In some cases  
we deliberately targeted some CEMP stars of extremely
low metallicity in order to gain some further insight
on the properties of these stars.

CEMP stars can be rich in heavy elements, which
are normally built by both the ``s-'' and  ``r- processes''.
In this case they are called CEMP-rs stars, although the nucleosynthetic
origin
of the neutron capture elements in these
stars is not very clear. It may in fact be 
more complex than a simple
superposition of ``normal'' s- and r-processes \citep[see][for a discussion]{masseron10}; 
if they are rich only in heavy elements built by the s-process
they are denominated CEMP-s.
CEMP stars with a normal pattern of the heavy elements are called CEMP-no stars
\citep[see e.g.][]{BC05}. 
The prototypical CEMP-no star is CS\,22957-027 (\citealt{norris97,bonifacio98},
with $[{\rm Fe/H}]=-3.43$, A(C)=7.13).
It is difficult to derive the chemical content of heavy elements in unevolved 
stars, and even more in the case of low- and medium-resolution spectra.
In the present sample two stars can be classified as 
CEMP-rs (SDSS\,J1137+2553) and CEMP-s (SDSS\,J1245-0738) on the basis of their
Ba abundance, following \citet{masseron10}.

An interesting issue is the lithium abundance in unevolved
CEMP stars compared to carbon-normal unevolved stars.
\citet{spite82} discovered that the warm metal-poor unevolved stars
share a common lithium abundance, independent of effective
temperature or metallicity.  This is referred to
as the {\em Spite plateau}. The unevolved CEMP stars usually display
lithium abundances that are lower than the  {\em Spite plateau},
although some of them do lie on the  {\em Spite plateau}.
Given that lithium is destroyed at temperatures higher
than $2.5\times 10^6$ K, the presence or absence of lithium
and its precise abundance place strong constraints on the
temperatures experienced by  the material present in the
stellar atmospheres and on possible lithium production
events. 

The bulk of the metals observed in stars has been 
produced by supernovae. The properties of
CEMP stars can provide evidence of the existence and
possible role of faint supernovae.
A faint supernova is a core-collapse supernova such that its kinetic
energy is of the order of a few $10^{50}$ erg, the velocity
of the ejecta $\le 1000$ \kms, and the mass
of $^{56}$Ni of the order of a few $10^{-3}$ M$_\odot$.
In such an explosion only the upper layers, 
rich in lighter elements (up to magnesium), are ejected and
recycled in the interstellar medium. The deeper layers, rich
in iron and heavier elements, fall back onto the compact
object and do not contribute to the enrichment of the 
interstellar medium \citep[see e.g.][and references therein]{ishigaki}.

In this paper we present abundance determinations for
six extremely iron-poor CEMP stars of which four are
studied here for the first time.
SDSS\,J1742+2531 was already analysed on the basis
of an X-Shooter spectrum by  \citet{gto13}. 
In that paper we were able to provide the abundance of
C, from the G-band and of Ca, from the \ion{Ca}{ii}
K line. No lines of Fe, or any other element
were detectable on that spectrum.
In this paper we give an account of our further observations
of this star conducted with UVES and providing higher
resolution and S/N spectra. 
Star SDSS\,J1245-0738 has already been studied by \citet{aoki13}.

%%%%%%%%%%%%%%%%%%%%%%%%%%%%%%%%%%%%%%%%%%%%%%%%%%%%%%%%%%%%%%%%%%
\section{Observations and data reduction}

The coordinates and SDSS photometry of our programme stars
are summarised in Table \ref{allstar}.

\subsection{SDSS\,J0212+0137}
For the star SDSS\,J0212+0137 we have nine 3000\,s and two 3700\,s exposures
taken with UVES
in the  437+760\,nm setting, collected between August 6 and August 31, 2013. 
The  slit width was set to 1\farcs{4}
to minimise the light losses, and the 
binning $2\times 2$. In this case the resolving power was set by
the seeing and since most observations were taken with a seeing around 1$''$
on the coadded spectrum, it is of the order of 47\,000.
The spectra were reduced
in the standard way
using the UVES pipeline \citep{ballester}
within the ESO Common Pipeline Library. 
The radial velocity was measured on each individual spectrum and then
spectra obtained on the same day were coadded and the radial
velocity measured from this coadded spectrum.
In both cases it was apparent that the radial velocity  
 varied nearly linearly 
over the observation period. 
Both the radial velocities from the individual spectra
and the daily means are given in Table \ref{RV}. 
The line-to-line scatter in the velocities of each
individual spectrum was typically less than 1 \kms and  
always less than 1.4 \kms.
For comparison the SDSS radial velocity
for this star is --2.7$\pm 3$\kms,  the spectrum was taken
on the 27-09-2009, MJD 5502. Further monitoring of the radial velocity
of this star is encouraged. 
The coadded spectrum has S/N = 79 at 392\,nm, 87 at 450\,nm,
134 at 670\,nm, and 210 at 850\,nm.

\subsection{SDSS\,J0929+0238}
\begin{figure}
\begin{center}
\resizebox{\hsize}{!}{\includegraphics[clip=true]{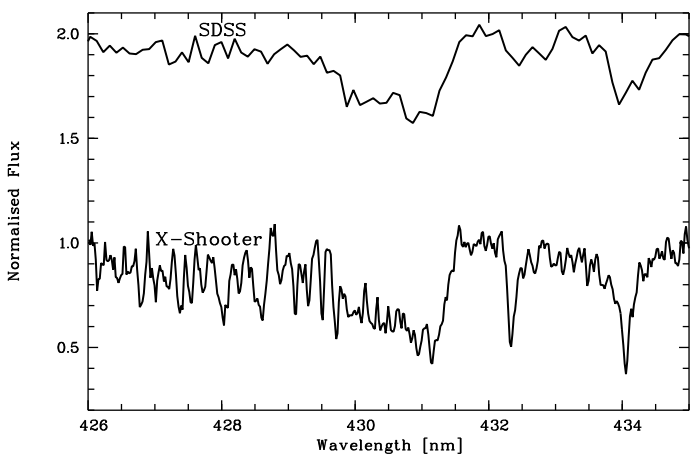}}
\end{center}
\caption[]{SDSS\,J0929+0238, the G-band in the SDSS and X-Shooter spectra.
\label{hydragband}}
\end{figure}

SDSS\,J0929+0238 was  observed with X-Shooter 
on May 12, 2013,
for an exposure time of
3000\,s in the IFU mode.
The S/N is  about 20 at 400\,nm.
Both the X-Shooter spectrum and the low-resolution spectrum of SDSS show an obvious G-band  (see Fig.\,\ref{hydragband}),
and the star was targeted for this reason as well as for the very weak \ion{Ca}{ii} K line
(see Fig.\,\ref{hydra_leo2}). 
This line and the G-band are the only metallic features we can detect
from the X-Shooter spectrum; they  provide only upper limits for iron
and other elements. 
The radial velocity as measured from the  \ion{Ca}{ii} K line
is +398 \kms \ with an error estimate of 10 \kms.
This is considerably lower than the velocity estimate of
SDSS (+467 \kms $\pm$ 10 \kms). However, considering that
the SDSS measurement can rely only on the Balmer lines,  
unsuitable for accurate radial velocity measurements owing to their breadth 
(see e.g. \citealt{jonay08}), it may well be that the error
in the SDSS measurement is much larger than its formal estimate  
and thus the two radial velocities are consistent.
It would  nevertheless be interesting to monitor the radial
velocity of this star since the difference between our measurement
and the SDSS may indicate the presence of radial velocity variations.

\subsection{SDSS\,J1035+0641}
SDSS\,J1035+0641   was observed  with X-Shooter 
on April 30 and May 1, 2013,
on each occasion for an exposure time of 3000\,s in the IFU mode.
The S/N of the coadded spectrum 
is about 35 at 400\,nm.
No  metallic features are detectable in our spectrum 
except for the  \ion{Ca}{ii} K and H lines
and the G-band.
The radial velocities measured from the \ion{Ca}{ii} K line
are -85 \kms and -70 \kms, with an estimated error of 10 \kms, 
thus making the two radial velocities consistent within $1\sigma$.
The SDSS radial velocity estimate is -47 \kms $\pm$ 6 \kms , 
the same caveats apply here as discussed for 
SDSS\,J0929+0238, and we  cannot be sure that this difference
is real.

We also observed the star with UVES in service mode
between February and April 2014 \relax in the standard setting
390+580\,nm. The slit was set to 1\farcs{4} and the CCD binned
$2\times2$, providing a resolving power $R \sim 32\,000$.
Eleven observing blocks, each for an exposure of 3005\,s,  which were executed
for a total time on target of about 9.2 hours.
The reduced data was retrieved from the ESO archive. 
The median S/N of the individual spectra ranges from 3 to 6 \relax in
the blue. This makes it very difficult to measure the radial velocities
on the individual spectra since we have to rely only on the Balmer lines.
These rough measurements showed that the radial velocity of the different
spectra was constant to within $\pm 5$ \kms. We therefore decided to coadd
the spectra, after applying the appropriate barycentric correction to 
each one. The S/N ratio of the coadded spectrum is about 19 at 400\,nm
and 40 at 670\,nm. The radial velocity measured on the coadded
spectrum is -70.2 \kms with an estimated error of 2 \kms,
which is consistent with the value  from  the X-Shooter spectra   
taken in 2013.

\subsection{SDSS\,J1137+2553}
For star SDSS\,J1137+2553 we have a single UVES
spectrum of 3005 s observed on July 2, 2013, with the
same set-up  used  for the observations
of SDSS\,J0212+0137. The seeing was 1\farcs{1}.
The spectrum has S/N=30 at 405\,nm.
The radial velocity measured from this spectrum is
$137.7\pm 0.4$ \kms in excellent agreement with the
value from the SDSS spectrum ($138.8\pm 3$ \kms ).

\subsection{SDSS\,J1245-0738}
For star SDSS\,J1245-0738 we have a single UVES spectrum
of 972\,s observed on May 9, 2013, with the same setup
used for SDSS\,J0212+0137 and SDSS\,J1137+2553.
The seeing was 1\farcs{2}. The spectrum is of poor
quality and attains S/N=8 at 405\,nm. In order to analyse
it we smoothed it by rebinning it by a factor of 3 and this
was sufficient to allow us to measure several metallic lines.
The radial velocity measured from this spectrum is $77.7\pm 1.27$ \kms,
in excellent agreement with the measurement from the SDSS spectrum
($75\pm 4$ \kms).
The star was also  studied by \citet{aoki13} using a 20 minutes
Subaru spectrum taken on March 10, 2008. Their measured radial
velocity was +76.88 $\pm 0.27$ \kms; this is again in agreement
with our measured radial velocity.

\begin{table}
\caption{\label{RV} Barycentric radial velocities for SDSS\,J0212+0137 }
\centering
\begin{tabular}{lll}
\hline\hline
Date  & $V_{bary}$ & JD - 2400000.5 \\
      & \kms       & days  \\ 
\hline
2013-08-06-01.279&2.39&56510.378487\\
2013-08-06-04.208&3.44&56510.271576\\
2013-08-06-22.838&3.67&56510.307209\\
2013-08-06-43.988&3.48&56510.342870\\
2013-08-12-13.675&3.87&56516.372381\\
2013-08-12-36.536&3.26&56516.253895\\
2013-08-12-46.360&3.36&56516.298453\\
2013-08-12-52.547&2.77&56516.335330\\
2013-08-13-10.375&3.71&56517.369565\\
2013-08-31-18.454&3.98&56535.273130\\
2013-08-31-39.166&5.73&56535.235870\\
\hline
\multicolumn{3}{c}{Daily mean velocities}\\
\hline
2013-08-06&$3.36\pm 0.53$&56510.325000\\
2013-08-12&$3.48\pm 0.52$&56516.315000\\
2013-08-13&$3.71\pm 0.55$&56517.369565\\
2013-08-31&$5.32\pm 0.65$&56535.118070\\
\hline
\end{tabular}
\end{table}

%%% FIGURE %%%%%%%%%%%%%%%%
%%% FIGURE %%%%%%%%%%%%%%%%

\subsection{SDSS\,J1742+2531}
SDSS\,J1742+2531  
was observed with UVES at the VLT-Kueyen telescope
in service mode, between April 17 and August 6, 2013.
Altogether a total of 29 observing blocks of 3000\,s exposure each were executed,
14 \relax in the setting 390+580\,nm and 15 \relax in the 
setting 437+760\,nm. One of the frames in the 390+580\,nm
setting had no useful signal. Our analysis is therefore 
based on the remaining 28  observing blocks. 
The projected slit on the sky was 1\farcs{0}, and the CCD
was binned $1\times 1$,
providing a resolving
power of about 47\,000.
The spectra have been reduced using the UVES pipeline 
in the same way as those of 
SDSS\,J0212+0137.
After performing the barycentric corrections all the individual spectra appeared to be
at the same radial velocity, to better than 0.5\,\kms. We therefore
coadded all the 28 spectra in the intervals common to the two
settings:  376-452\,nm and 583-680\,nm.
In the range 328-376\,nm we coadded the 13 
spectra of the setting 390+580\,nm and in the range
680-946\,nm the spectra of the setting 437+760\,nm.
The S/N ratio of the coadded spectra is 3 at 336\,nm,
20 at 382\,nm, 23 at 450\,nm, 44 at 518\,nm, 45 at 670\,nm
and 36 at 850\,nm.
The measurement of the radial velocity is not trivial with so few
detected lines. We obtained a first estimate of the radial velocity
from the cores of the Balmer lines, and then refined it using
the measured central wavelength of the seven detected metallic lines,
the resulting radial velocity is $-207.5$ \kms$\pm 0.5$ \kms (r.m.s.).
This agrees with the SDSS radial velocity ($-221$ \kms $\pm 11$ \kms) 
within 1.2$\sigma$.

%%% FIGURE %%%%%%%%%%%%%%%%
\begin{figure}
\begin{center}
\resizebox{\hsize}{!}{\includegraphics[clip=true]{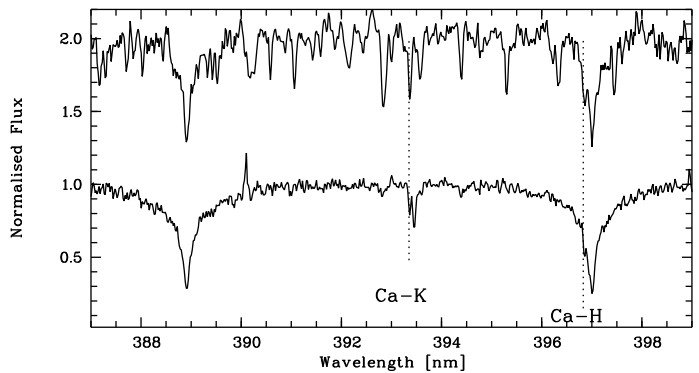}}
\end{center}
\caption[]{The X-Shooter spectra around the \ion{Ca}{ii} H \& K lines
of SDSS\,J0929+0238 (the spectrum is vertically
displaced by 1 unit for display purposes) and SDSS\,J1035+0641.
The dotted lines mark the position of the stellar
\ion{Ca}{ii} H \&K lines. 
The CH lines are much more prominent in the former because the carbon
abundance is almost 1\,dex larger and the effective temperature
is almost 300\,K cooler.}
\label{hydra_leo2}
\end{figure}
%%% FIGURE %%%%%%%%%%%%%%%%

%%% FIGURE %%%%%%%%%%%%%%%%
\begin{figure}
\begin{center}
\resizebox{\hsize}{!}{\includegraphics[clip=true]{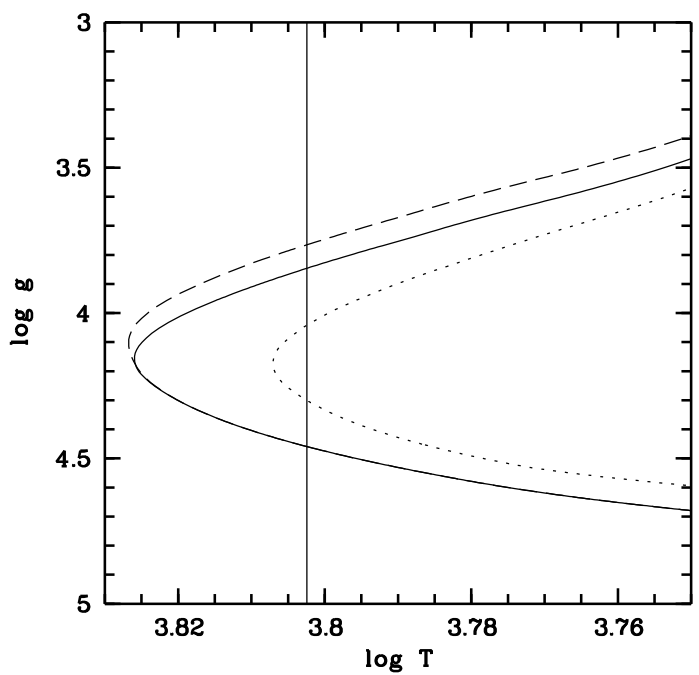}}
\end{center}
\caption[]{
Isochrones computed with the FRANEC code 
for 14 Gyrs and $Z=2\times 10^{-6}$ (dashed line),
$Z=2\times 10^{-5}$ (solid line)
and $Z=2\times 10^{-4}$ (dotted line).
The vertical line marks our adopted \teff for SDSS\,J1742+2531.
\label{evol}}
\end{figure}
%%% FIGURE %%%%%%%%%%%%%%%%

%%% FIGURE %%%%%%%%%%%%%%%%
\begin{figure}
\begin{center}
\resizebox{\hsize}{!}{\includegraphics[clip=true]{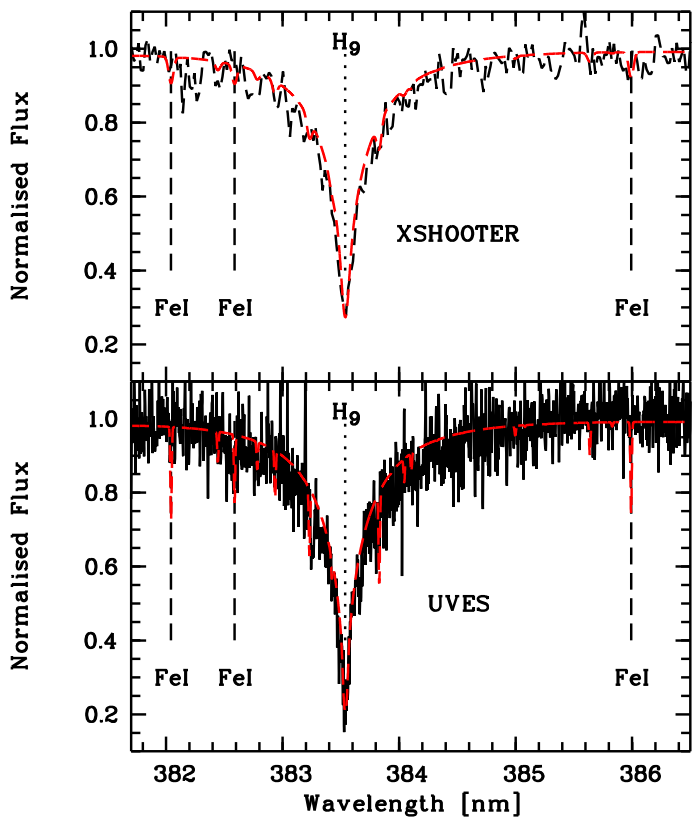}}
\end{center}
\caption[]{SDSS\,J1035+06411: the X-Shooter (top panel) and 
UVES (bottom panel) spectra in the region covering
the three strongest \ion{Fe}{i} lines. The comparison spectrum is a synthetic
spectrum computed with [Fe/H]=--4.5.
\label{leo2Fe}}
\end{figure}
%%% FIGURE %%%%%%%%%%%%%%%%

%%%%%%%%%%%%%%%%%%%%%%%%%%%%%%%%%%%%%%%%%%%%%%%%%%%%%%%%%%%%%%%%%%

\section{Analysis and results}

\begin{table}
\renewcommand{\tabcolsep}{1pt}
\caption{\label{atm} Atmospheric parameters }
\centering
\begin{tabular}{lcccrrr}
\hline
SDSS ID          &\teff & log g$^a$ & $\xi^b$ & [Fe/H] & [Ca/H] & A(C) \\
                 &  K   & c.g.s  & \kms &\\
\hline\hline
SDSS\,J0212+0137 & 6333 & 4.0 &1.3 & $-3.59$  & $-2.81$ &  7.12\\
SDSS\,J0929+0238 & 5894 & 3.7 &1.5 & $<-3.81$ & $-4.02$ &  7.70\\
SDSS\,J1035+0641 & 6262 & 4.0 &1.5 & $<-5.07$ & $-5.00$ &  6.90 \\
SDSS\,J1137+2553 & 6310 & 3.2 &1.5 & $-2.70$  & $-2.18$ & 8.60 \\ 
SDSS\,J1245-0738 & 6110 & 2.5 & 3.0 & $-3.21$ & $-2.35$ & 8.65 \\   
SDSS\,J1742+2531 & 6345 & 4.0 & 1.5 & $-4.80$  & $-4.56$ &  7.26 \\
\hline
\multicolumn{7}{l}{$^a$ logarithm of the gravitational acceleration at the surface of}\\
\multicolumn{7}{l}{\phantom{$^a$} the stars expressed in cm s$^{-2}$}\\
\multicolumn{7}{l}{$^b$ microturbulent velocity}\\
\hline
\end{tabular}
\end{table}

\subsection{Atmospheric parameters\label{param}}

The adopted atmospheric
parameters for the programme stars are summarised in Table \ref{atm}.

\subsubsection{SDSS\,J0212+0137}
For SDSS\,J0212+0137, we derived the temperature of 6333\,K from the $(g-z)$ colour \cite{Ludwig08},
with an extinction of E($g-z$)=0.07.
The adopted gravity was  \logg =4.0, that achieves the iron
ionisation equilibrium, although only one \ion{Fe}{ii} line could be measured.
The effective temperature of this star is very near to that of SDSS\,J1742+2531, discussed later
in this section,  
as is its global metallicity, dominated by the carbon (and presumed oxygen)
overabundance with respect to iron. 
Thus the same isochrone
is representative both for SDSS\,J0212+0137 and
 SDSS\,J1742+2531,  and \logg = 4.0 is 
appropriate for this \teff . 
For this star we exclude an RR Lyr nature; the 
observed radial velocity variations are not large enough and
in any case we do not expect to observe a nearly
linear variation of the radial velocity as shown in Table \ref{RV}.

\subsubsection{SDSS\,J0929+0238}
 
For SDSS\,J0929+0238 we adopted a temperature of 5894\,K, derived from
the $(g-z)$ calibration. This temperature involves a gravity from isochrones
of \logg =4.5 or 3.7. We prefer the lower value of gravity as it provides a temperature from the
wings of H$\alpha$ consistent with the photometric value.

\subsubsection{SDSS\,J1035+0641}
For SDSS\,J1035+0641 we adopted a temperature of 6262\,K, derived from
the $(g-z)$ calibration. 
We used the same FRANEC isochrones described below
in Sect. \ref{ercolinapar} to estimate the surface gravity,
given the effective temperature.
If the star is slightly evolved
it has log g = 4.0, while if it is still on the main sequence
it should have log g = 4.4.

\subsubsection{ SDSS\,J1137+2553 and SDSS\,J1245-0738}
Also for SDSS\,J1137+2553 and SDSS\,J1245-0738 we adopted the
effective temperature derived from the $g-z$ calibration that
appears to be consistent with the wings of H$\alpha$ and the
iron excitation equilibrium.
The gravity  was derived by requiring the iron ionisation balance,
to within the available uncertainty. 
These  imply that SDSS\,J1137+2553 is a subgiant star
and SDSS\,J1245-0738 is an HB star. The latter should be further
monitored photometrically to check if it is an RR-Lyr.
In the case it were an RR-Lyr our photometric temperature may be
inadequate to describe our available spectrum.
We note however that the lack of any radial velocity variation
between our observation, that of \citet{aoki13} and the SDSS
does not support an RR-Lyr status. Our
adopted atmospheric temperature is very close to that
adopted by \citet[][\teff = 6108 K]{aoki13} while we differ on the surface gravity.
\citet{aoki13} {\em assumed} their stars to be 
turn-off without requiring the iron ionisation balance.

\subsubsection{SDSS\,J1742+2531\label{ercolinapar}}
For SDSS\,J1742+2531, in \citet{gto13} we adopted an effective
temperature of 6345\,K deduced from the $(g-z)_0$ colour
and consistent with the wings of H$\alpha$ from our
X-Shooter spectrum. The wings of  H$\alpha$
from the UVES spectrum are also consistent with this
temperature. We adopted log g = 4.0 for the surface
gravity, that is appropriate for a metal-poor TO star
of this effective temperature.
In Fig.\,\ref{evol} we show a comparison with two
isochrones computed with the FRANEC code (Chieffi, Limongi - private communication) 
for 14 Gyr, $Z=2\times 10^{-4}$ (dotted line), $Z=2\times 10^{-5}$ (solid line) and 
$Z=2\times 10^{-6}$ (dashed line). 
In spite of its very low iron abundance 
($[{\rm Fe/H}]=-4.8$, see Table \ref{abund}), due to its large
carbon enhancement,
SDSS\,J1742+2531 should have a metallicity $1.8\times 10^{-4} \le Z < 2.8 \times 10^{-4}$,
depending on its actual oxygen abundance, where the upper limit is given in
Table \ref{abund} and the lower limit corresponds
to $[{\rm O/Fe}]=+0.4$. Taking the isochrone   $Z=2\times 10^{-4}$  as reference, 
we have two possible solutions for the surface gravity for our assumed \teff :
\logg = 4.0, if the star is evolved past the turn off (TO), or
log g = 4.3 if the star has not yet reached the TO.
In the absence of any reliable gravity indicator we assume
log g = 4.0, since this implies a larger distance.
Given the rarity of these stars we assume that it is more likely
that we found one of the brighter members of this population.
In the next section, we nevertheless provide also the abundances
computed assuming the higher gravity.

A possible concern is that the star may be indeed a horizontal branch
star. In this case, it could be an RR-Lyr. 
To check if the star is an RR Lyr we measured the radial
velocities of the \ion{Ca}{ii} H and K lines on all our available
spectra and we did not find any variations above 1\kms.
The H$\alpha$ line is also clearly detected in the individual
spectra, yet given its broad nature the accuracy
of the radial velocities determined is inferior to
that of the metallic lines. 
We  inspected the H$\alpha$ line in all the spectra
and found no evidence of  variations above 5\,\kms nor
of the presence of P-Cyg profiles.
The radial velocity amplitudes expected for an RR Lyr star are
larger than 50 \kms\ for metallic lines and
larger than 100 \kms\ for H$\alpha$ 
\citep{rrlyr}.
The typical periods of RR Lyr stars are less than a day and
they should be adequately sampled by our data set.
We obtained a photometric series in the $r$ band of about 2h length
with the AFOSC camera at the 1.82m telescope of
Cima Ekar (Asiago), and the magnitude of this star seems
constant with an r.m.s. of 0.02 mag, compatible with the S/N
of the images.
We are obtaining also longer
photometric series with the Schmidt telescope
of the Asiago Observatory 
in the $r$ band to check for variability on longer time scales. 
All this photometric data will be fully discussed elsewhere.
At the time of writing, we have no evidence
of RR-Lyr -- like variability of this star.

By looking at the figure 1 of \citet{sandage} and estimating
the (B-V)$_0$ of the star to be 0.39, from its $(g-r)_0$ 
colour\footnote{From Table 1 of \citet{jester}: B-V $= 0.98\times (g-r)+0.22$}
we see that, as long as the absolute magnitude of the star is less than
roughly 0.8, it lies in the instability strip.
The isochrone shown in Fig.\,\ref{evol} implies also that
a subgiant star with log g = 3.3 has  M$_V$=2.7, thus 
keeping it well out of the instability strip. Horizontal branch
stars, of a temperature
similar to  that of SDSS\,J1742+2531, should have a gravity that is this
low or even lower (see figure 3 of \citealt{camilla} using
the evolutionary tracks  of \citealt{cassisi}), depending on the star's mass.
Moreover as discussed below in Sect.\,\ref{ercolinachem}, a gravity as low as log g = 3.3, 
would bring about a 0.2\,dex discrepancy between the \ion{Ca}{ii} K line
and the \ion{Ca}{ii} triplet, and lower surface gravities would increase
this discrepancy.

\subsection{Chemical Abundances}

To derive abundances and upper limits either from the equivalent
widths or using synthetic spectra and line-fitting
we used {\tt turbospectrum} \citep{alvarez_plez,2012ascl.soft05004P}.
We based our analysis 
on MARCS 1D LTE  (Local Thermodynamic Equilibrium) 
model atmosphere \citep{G2008}, that were interpolated in the grid
computed for the TOPoS survey \citep{topos1} 
for the atmospheric parameters  of each star.
As solar abundances we adopted  carbon, nitrogen, 
oxygen and iron from \citet{abbosun},
for the other elements \citet{lodders09}.
If too few iron lines were measured to allow the
determination of the microturbulence 
we adopted 1.5\,\kms as in \citet{gto13} 

\subsubsection{SDSS\,J0212+0137}

The  higher abundance
in iron and other elements in SDSS\,J0212+0137, coupled
with the good quality of the spectrum 
allows us  to obtain an extensive 
chemical inventory. The 
line by line abundances are provided in Table \ref{linescemp}
and the mean abundances are provided in
Table \ref{abucemp}.
The Fe abundance is derived from 17 \ion{Fe}{i} lines
and of the 8 measured \ion{Fe}{ii} lines we retain
only one, that has an equivalent width larger
than 0.5\,pm\footnote{$1 {\rm pm}=10^{-12} {\rm m}$} and this line provides the same abundance
as the mean of the 17 \ion{Fe}{i} abundances for our adopted surface gravity
log g = 3.7.
We measure both Sr and Ba in this star and their abundance
classifies unmistakably the star as a CEMP-no.
Remarkably we measure also the lithium abundance in this star
that places it on the {\em Spite plateau} \citep{spite82}. This pattern is indeed  similar
to that of SDSS\, J1036+1212 \citep{behara}, that is a CEMP-no/s
\footnote{A CEMP-no/s is a star with +0.5 
$<$ [Ba/Fe] $<$ +1.0 \citep{sivarani}.}, 
has a relatively low-carbon abundance
and has a lithium abundance in agreement with the {\em Spite plateau}.

\begin{longtab}
\setlength{\tabcolsep}{4pt}
\begin{longtable}{lllrllllll}
\caption{\label{linescemp} Line-by-line abundances of SDSS\,J0212+0137, 
SDSS\,J1137+2553,  and SDSS\,J1245-0738
}\\
\hline\hline
\multicolumn{4}{c}{}&
\multicolumn{2}{c}{SDSS\,J0212+0137}&
\multicolumn{2}{c}{SDSS\,J1137+2553}&
\multicolumn{2}{c}{SDSS\,J1245-0738}\\
Ion          & $\lambda^a$ &$\chi^b$ & log gf$^c$  & EW$^d$   & A(X)$^e$& EW$^d$& A(X)$^e$&EW$^d$& A(X)$^e$ \\
             & nm        & eV    &         & pm   &      &         \\
\hline\hline
\endfirsthead
\caption{continued.}\\
\hline\hline
\multicolumn{4}{c}{}&
\multicolumn{2}{c}{SDSS\,J0212+0137}&
\multicolumn{2}{c}{SDSS\,J1137+2553}&
\multicolumn{2}{c}{SDSS\,J1245-0738}\\
Ion          & $\lambda^a$ &$\chi^b$ & log gf$^c$  & EW$^d$   & A(X)$^e$& EW$^d$   & A(X)$^e$& EW$^d$   & A(X)$^e$ \\
             & nm        & eV    &         & pm   &      &         \\
\hline\hline
%       lambda    Exc  log(gf)           obs.eqw+- error   delta    low      abund    up
\hline
\endhead
\hline
\endfoot
%
%                                 SDSS J0212+0137  SDSS J1137+2553  SDSS J1245-0738
%       lambda    Exc  log(gf)     EW    abund      EW    abund      EW    abund
\ion{Li}{i} & 670.7761& 0.00&$-0.009$&   synth&$  2.08$&    synth&$  2.30$&         &$      $\\
\ion{Li}{i} & 670.7912& 0.00&$-0.299$&   synth&$  2.08$&    synth&$  2.30$&         &$      $\\
\ion{O} {i} & 777.1941& 9.15&$ 0.369$&   synth&$  6.70$&    synth&$  7.18$&    synth&$  7.30$\\
\ion{O} {i} & 777.4161& 9.15&$ 0.223$&   synth&$  6.70$&    synth&$  7.18$&    synth&$  7.30$\\
\ion{O} {i} & 777.5388& 9.15&$ 0.001$&   synth&$  6.70$&    synth&$  7.18$&    synth&$  7.30$\\
\ion{Na}{i} & 588.9951& 0.00&$ 0.117$&        &$      $&    synth&$  6.10$&    synth&$  4.00$\\
\ion{Na}{i} & 589.5924& 0.00&$-0.184$&        &$      $&    synth&$  6.30$&    synth&$  4.50$\\
\ion{Mg}{i} & 382.9355& 2.71&$-0.231$&   synth&$  4.45$&    synth&$  6.17$&    synth&$  5.30$\\
\ion{Mg}{i} & 383.2304& 2.71&$ 0.146$&   synth&$  4.55$&    synth&$  6.17$&         &$      $\\
\ion{Mg}{i} & 383.8290& 2.72&$ 0.415$&   synth&$  4.60$&    synth&$  6.17$&    synth&$  4.80$\\
\ion{Al}{i} & 394.4006& 0.00&$-0.623$&   synth&$  2.34$&         &$  3.30$&         &$      $\\
\ion{Al}{i} & 396.1520& 0.01&$-0.323$&   synth&$  2.28$&     7.14&$  3.30$&     8.47&$  3.10$\\
\ion{Si}{i} & 390.5523& 1.91&$-1.041$&   synth&$ 4.15 $&    synth&$  5.10$&    synth&$  4.30$\\
\ion{Ca}{i} & 422.6728& 0.00&$ 0.244$&    7.86&$  3.35$&    11.59&$  4.24$&    synth&$  3.05$\\
\ion{Ca}{i} & 445.4779& 1.90&$ 0.258$&    1.06&$  3.36$&     4.17&$  4.17$&    synth&$  4.00$\\
\ion{Ca}{ii} & 393.3663& 0.00&$ 0.105$&   synth&$  3.50$&    synth&$  4.15$&    synth&$  3.97$\\
\ion{Ca}{ii} & 396.8469& 0.00&$-0.200$&   synth&$  3.55$&    synth&$  4.15$&    synth&$  4.00$\\
\ion{Ca}{ii} & 849.8023& 1.69&$-1.416$&   synth&$  3.80$&    synth&$  4.60$&    synth&$  4.20$\\
\ion{Ca}{ii} & 854.2091& 1.70&$-0.463$&   synth&$  4.00$&         &$      $&         &$      $\\
\ion{Ca}{ii} & 866.2170& 1.69&$-0.723$&   synth&$  4.00$&         &$      $&         &$      $\\
\ion{Sc}{ii} & 424.6822& 0.31&$ 0.242$&    1.48&$  0.00$&     7.19&$  1.03$&     6.04&$  0.16$\\
\ion{Ti}{ii} & 375.9291& 0.61&$ 0.280$&        &$      $&    10.42&$  2.72$&    12.92&$  1.90$\\
\ion{Ti}{ii} & 376.1320& 0.57&$ 0.180$&    7.02&$  2.13$&     9.74&$  2.60$&    14.24&$  2.24$\\
\ion{Ti}{ii} & 391.3461& 1.12&$-0.420$&    2.34&$  2.08$&     5.29&$  2.43$&     5.71&$  1.96$\\
\ion{Ti}{ii} & 441.7714& 1.16&$-1.190$&     .56&$  2.11$&     2.23&$  2.54$&         &$      $\\
\ion{Ti}{ii} & 444.3794& 1.08&$-0.720$&     .82&$  1.74$&     3.45&$  2.27$&     5.09&$  2.09$\\
\ion{Ti}{ii} & 444.4555& 1.12&$-2.240$&        &$      $&     0.72&$  2.97$&         &$      $\\
\ion{Ti}{ii} & 445.0482& 1.08&$-1.520$&        &$      $&         &$      $&         &$      $\\
\ion{Ti}{ii} & 446.8507& 1.13&$-0.600$&    1.63&$  2.01$&     4.41&$  2.38$&         &$      $\\
\ion{Ti}{ii} & 450.1270& 1.12&$-0.770$&    1.30&$  2.05$&     3.32&$  2.32$&         &$      $\\
\ion{Ti}{ii} & 453.3960& 1.24&$-0.530$&    1.23&$  1.89$&     4.70&$  2.46$&     8.10&$  2.42$\\
\ion{Ti}{ii} & 456.3757& 1.22&$-0.690$&    1.13&$  1.99$&     3.20&$  2.31$&     6.46&$  2.36$\\
\ion{Ti}{ii} & 457.1968& 1.57&$-0.320$&    1.04&$  1.90$&     3.64&$  2.35$&     2.45&$  1.72$\\
\ion{Cr}{i} & 425.4336& 0.00&$-0.114$&    1.32&$  2.09$&     5.24&$  3.03$&     2.41&$  2.17$\\
\ion{Fe}{i} & 375.8233& 0.96&$-0.027$&        &$      $&     9.88&$  5.00$&         &$      $\\
\ion{Fe}{i} & 376.3789& 0.99&$-0.238$&    4.93&$  3.85$&     9.39&$  5.10$&    11.51&$  4.37$\\
\ion{Fe}{i} & 376.5539& 3.24&$ 0.482$&        &$      $&     3.55&$  4.78$&         &$      $\\
\ion{Fe}{i} & 376.7192& 1.01&$-0.389$&    4.05&$  3.81$&     7.02&$  4.52$&     7.03&$  3.84$\\
\ion{Fe}{i} & 378.7880& 1.01&$-0.859$&    2.78&$  3.97$&     6.37&$  4.79$&    12.08&$  5.11$\\
\ion{Fe}{i} & 380.5342& 3.30&$ 0.312$&        &$      $&         &$      $&         &$      $\\
\ion{Fe}{i} & 381.5840& 1.49&$ 0.237$&    5.27&$  3.90$&     8.90&$  4.88$&     7.04&$  3.66$\\
\ion{Fe}{i} & 382.0425& 0.86&$ 0.119$&    7.42&$  4.07$&    10.31&$  4.84$&     7.64&$  3.27$\\
\ion{Fe}{i} & 382.4444& 0.00&$-1.362$&    4.84&$  4.04$&     8.39&$  5.02$&     6.94&$  3.84$\\
\ion{Fe}{i} & 382.5881& 0.92&$-0.037$&    5.98&$  3.85$&         &$      $&         &$      $\\
\ion{Fe}{i} & 382.7822& 1.56&$ 0.062$&    4.27&$  3.89$&     7.57&$  4.71$&    13.13&$  4.90$\\
\ion{Fe}{i} & 384.0437& 0.99&$-0.506$&    3.59&$  3.79$&     8.40&$  5.02$&         &$      $\\
\ion{Fe}{i} & 384.9966& 1.01&$-0.871$&    2.39&$  3.88$&     7.75&$  5.20$&    12.95&$  5.27$\\
\ion{Fe}{i} & 385.0818& 0.99&$-1.734$&    0.89&$  4.19$&     4.04&$  5.08$&         &$      $\\
\ion{Fe}{i} & 385.6371& 0.05&$-1.286$&    5.04&$  4.06$&     8.75&$  5.11$&     9.38&$  4.15$\\
\ion{Fe}{i} & 385.9911& 0.00&$-0.710$&    7.67&$  4.22$&     9.46&$  4.73$&     9.44&$  3.53$\\
\ion{Fe}{i} & 386.5523& 1.01&$-0.982$&    2.31&$  3.97$&     5.92&$  4.78$&     8.03&$  4.55$\\
\ion{Fe}{i} & 387.8018& 0.96&$-0.914$&    2.65&$  3.94$&     6.21&$  4.73$&         &$      $\\
\ion{Fe}{i} & 389.9707& 0.09&$-1.531$&    4.07&$  4.09$&     7.06&$  4.81$&         &$      $\\
\ion{Fe}{i} & 392.0258& 0.12&$-1.746$&    2.26&$  3.90$&     6.24&$  4.81$&     3.18&$  3.80$\\
\ion{Fe}{i} & 392.2912& 0.05&$-1.651$&    2.93&$  3.91$&     8.08&$  5.22$&     9.55&$  4.53$\\
\ion{Fe}{i} & 400.5242& 1.56&$-0.610$&    2.41&$  4.10$&     4.65&$  4.58$&     2.83&$  3.94$\\
\ion{Fe}{i} & 404.5812& 1.49&$ 0.280$&    5.70&$  3.93$&     8.87&$  4.78$&     7.61&$  3.66$\\
\ion{Fe}{i} & 406.3594& 1.56&$ 0.062$&    4.48&$  3.91$&     8.04&$  4.80$&     9.94&$  4.27$\\
\ion{Fe}{i} & 407.1738& 1.61&$-0.022$&    3.39&$  3.79$&     7.75&$  4.84$&     8.22&$  4.15$\\
\ion{Fe}{i} & 413.2058& 1.61&$-0.675$&    1.09&$  3.76$&     5.55&$  4.88$&     4.03&$  4.24$\\
\ion{Fe}{i} & 414.3868& 1.56&$-0.511$&    1.93&$  3.86$&     5.23&$  4.60$&     8.86&$  4.67$\\
\ion{Fe}{i} & 415.6799& 2.83&$-0.809$&        &$      $&     1.20&$  5.03$&         &$      $\\
\ion{Fe}{i} & 418.7039& 2.45&$-0.548$&    0.51&$  4.02$&     2.22&$  4.76$&         &$      $\\
\ion{Fe}{i} & 418.7795& 2.42&$-0.554$&        &$      $&     2.68&$  4.86$&     2.79&$  4.66$\\
\ion{Fe}{i} & 419.1430& 2.47&$-0.666$&        &$      $&     1.32&$  4.61$&     2.43&$  4.74$\\
\ion{Fe}{i} & 419.9095& 3.05&$ 0.155$&    0.51&$  3.86$&     3.16&$  4.82$&     2.21&$  4.40$\\
\ion{Fe}{i} & 420.2029& 1.49&$-0.708$&    1.66&$  3.90$&     6.07&$  4.93$&     5.51&$  4.36$\\
\ion{Fe}{i} & 422.7427& 3.33&$ 0.266$&        &$      $&     3.05&$  4.94$&         &$      $\\
\ion{Fe}{i} & 425.0119& 2.47&$-0.405$&        &$      $&     2.60&$  4.73$&         &$      $\\
\ion{Fe}{i} & 426.0474& 2.40&$ 0.109$&    1.27&$  3.76$&     5.92&$  4.88$&     2.52&$  3.92$\\
\ion{Fe}{i} & 438.3545& 1.49&$ 0.200$&    5.55&$  3.93$&     8.97&$  4.82$&     8.39&$  3.81$\\
\ion{Fe}{i} & 440.4750& 1.56&$-0.142$&    3.52&$  3.86$&     7.08&$  4.67$&     9.01&$  4.30$\\
\ion{Fe}{i} & 444.3194& 2.86&$-1.043$&        &$      $&     0.73&$  5.03$&         &$      $\\
\ion{Fe}{i} & 446.6551& 2.83&$-0.600$&        &$      $&     1.96&$  5.07$&         &$      $\\
\ion{Fe}{i} & 452.8614& 2.18&$-0.822$&    0.83&$  4.26$&     2.54&$  4.85$&     2.86&$  4.69$\\
\ion{Fe}{i} & 489.1492& 2.85&$-0.112$&        &$      $&     3.52&$  4.95$&         &$      $\\
\ion{Fe}{i} & 490.3310& 2.88&$-0.926$&        &$      $&     0.97&$  5.05$&         &$      $\\
\ion{Fe}{i} & 491.8994& 2.87&$-0.342$&    0.61&$  4.23$&     1.30&$  4.59$&     1.34&$  4.45$\\
\ion{Fe}{i} & 492.0502& 2.83&$ 0.068$&    0.63&$  3.81$&     2.99&$  4.63$&     4.63&$  4.68$\\
\ion{Fe}{ii} & 423.3172& 2.58&$-1.947$&        &$      $&     2.65&$  4.60$&     3.08&$  4.33$\\
\ion{Fe}{ii} & 441.6830& 2.78&$-2.602$&        &$      $&     1.24&$  4.99$&         &$      $\\
\ion{Fe}{ii} & 449.1405& 2.86&$-2.756$&        &$      $&     0.25&$  4.46$&         &$      $\\
\ion{Fe}{ii} & 452.0224& 2.81&$-2.617$&        &$      $&     0.45&$  4.54$&         &$      $\\
\ion{Fe}{ii} & 454.1524& 2.86&$-2.973$&        &$      $&     0.46&$  4.95$&         &$      $\\
\ion{Fe}{ii} & 455.5893& 2.83&$-2.421$&        &$      $&     1.89&$  5.07$&     1.41&$  4.61$\\
\ion{Fe}{ii} & 492.3927& 2.89&$-1.504$&    0.65&$  3.93$&     4.98&$  4.90$&     4.15&$  4.31$\\
\ion{Co}{i} & 384.5461& 0.92&$ 0.010$&        &$      $&         &$      $&     3.00&$  2.64$\\
\ion{Co}{i} & 399.5302& 0.92&$-0.220$&        &$      $&     2.33&$  2.97$&         &$      $\\
\ion{Co}{i} & 412.1311& 0.92&$-0.320$&    0.78&$  2.49$&         &$      $&     2.83&$  2.91$\\
\ion{Ni}{i} & 380.7138& 0.42&$-1.205$&    0.90&$  2.82$&     2.11&$  3.25$&         &$      $\\
\ion{Ni}{i} & 385.8292& 0.42&$-0.936$&    1.27&$  2.72$&     4.10&$  3.44$&     5.48&$  3.29$\\
\ion{Sr}{ii} & 407.7709& 0.00&$ 0.167$&    2.58&$ -0.87$&    12.14&$  1.40$&    11.93&$ -0.17$\\
\ion{Sr}{ii} & 421.5519& 0.00&$-0.145$&    1.36&$ -0.96$&    11.64&$  1.60$&     7.26&$ -0.60$\\
\ion{Ba}{ii} & 413.0645& 2.72&$  hfs $&        &$      $&    synth&$  1.90$&         &$      $\\
\ion{Ba}{ii} & 455.4029& 0.00&$  hfs $&   synth&$ -1.40$&    synth&$  2.55$&    synth&$  0.35$\\
\ion{Ba}{ii} & 493.4829& 1.25&$  hfs $&        &$      $&    synth&$  2.55$&    synth&$  0.10$\\
\ion{Ba}{ii} & 585.3668& 0.60&$  hfs $&        &$      $&    synth&$  1.88$&         &$      $\\
\ion{Ba}{ii} & 614.1713& 0.70&$  hfs $&        &$      $&    synth&$  2.45$&         &$      $\\
\ion{Ba}{ii} & 649.6897& 0.60&$  hfs $&        &$      $&    synth&$  2.50$&         &$      $\\
\hline
\multicolumn{7}{l}{$^a$ Wavelength}  \\
\multicolumn{7}{l}{$^b$ Lower energy of the transition}  \\
\multicolumn{7}{l}{$^c$ Logarithm of the product of the oscillator strength of}  \\
\multicolumn{7}{l}{\phantom{$^c$} the transition and the statistical weight of the lower level}  \\
\multicolumn{7}{l}{$^d$ Equivalent width of the line}  \\
\multicolumn{7}{l}{$^e$ Abundance of the element A(X) = log (X/H) + 12 }  \\
\hline
\end{longtable}
\end{longtab}

\begin{table}
\caption{\label{abucemp} Mean abundances of SDSS\,J0212+0137}
\centering
\setlength{\tabcolsep}{2pt}
\begin{tabular}{lccrcc}
\hline\noalign{\smallskip}
Ion                & A(X)  &A(X)\sun& [X/H]     & $\sigma$ & N lines\\
\hline\hline\noalign{\smallskip}
\ion{Li}{i}        & 2.04   & 1.03   &          &      & 1\\
\ion{O}{i}         & 6.70   & 8.76   &   --2.06 &      & 3\\
\ion{Mg}{i}        & 4.53   & 7.54   &   --3.01 & 0.08 & 3\\
\ion{Al}{i}        & 2.31   & 6.47   &   --4.16 & 0.04 & 2\\
\ion{Si}{i}        & 4.15   & 7.52   &   --3.37 &      & 1\\
\ion{Ca}{i}        & 3.35   & 6.33   &   --2.98 &      & 1\\
\ion{Ca}{ii} H\&K & 3.52   & 6.33   &   --2.81 &      &  2\\
\ion{Ca}{ii} IR    & 3.93   & 6.33   &   --2.40 &      & 3\\
\ion{Sc}{ii}       & 0.00   & 3.10   &   --3.10 &      & 1\\
\ion{Ti}{ii}       & 1.99   & 4.90   &   --2.91 & 0.13 & 9\\
\ion{Cr}{i}        & 2.09   & 5.64   &   --3.55 &      & 1\\
\ion{Fe}{i}        & 3.93   & 7.52   &   --3.59 & 0.15 & 17\\
\ion{Co}{i}        & 2.49   & 4.92   &   --2.43 &      & 1\\
\ion{Ni}{i}        & 2.77   & 6.23   &   --3.46 & 0.07 & 2\\
\ion{Sr}{ii}       &--0.91  & 2.92   &   --3.83 & 0.06 & 2\\
\ion{Ba}{ii}       &--1.40  & 2.17   &   --3.57 &      & 1\\
\hline\noalign{\smallskip}
\multicolumn{6}{c}{Molecular bands}\\
\hline\noalign{\smallskip}
 C --CH (G-band)     & 7.12 & 8.50   &   --1.38 & & \\
 N --CN (388nm)      &$\le$6.90 & 7.86   & $\le$--0.96 & & \\
\hline\noalign{\smallskip}
\end{tabular}
\end{table}

\subsubsection{SDSS\,J0929+0238}
For SDSS\,J0929+0238 no metallic features could be safely  identified in the X-Shooter spectrum,
except for the \ion{Ca}{ii}-K line, which gives A(Ca)=2.3, and the G-band, that implies
a strong enhancement in carbon, A(C)=7.7.
A higher resolution spectrum is required to determine
the iron abundance of this star.
The C and Ca abundances as well as a few upper limits are
given in Table \ref{hydra}.

\begin{table}
\caption{\label{hydra} Abundances of SDSS\,J092912+023817}
\centering
\setlength{\tabcolsep}{1pt}
\begin{tabular}{lccrllll}
\hline\noalign{\smallskip}
\multicolumn{7}{c}{Atomic lines}  \\
\hline\noalign{\smallskip}
Ion & $\lambda$ &$\chi$ & log gf & EW & A(X) & A(X)$_{err}$& A(X)\sun\\
    & nm        & eV    &        & pm &      &             &      \\
\hline\noalign{\smallskip}
\ion{Li}{i} & 6707.761$^a$ &  0.00 &--0.009  & $<7.50 $& $<2.5^b$ &   & 1.03 \\
\ion{Mg}{i} & 518.3604 &  2.72 & --0.239 & $<7.50$ & $<4.10$ & & 7.54\\
\ion{Ca}{ii}& 393.3663 &  0.00 &  0.105  &  syn  & $\phantom{<}2.30$ & 0.07 & 6.33\\
\ion{Fe}{i} & 382.0425 &  0.86 &  0.119  &  $<2.73^c$ & $<2.38$ & & 7.52\\ 
\ion{Fe}{i} & 382.0425 &  0.86 &  0.119  &  $<8.91^d$& $<3.71$  & & 7.52\\ 
\ion{Sr}{ii}& 407.7709 &  0.00 &  0.167  &  $<7.80$ & $<0.04$ &  & 2.92\\
\ion{Ba}{ii}& 4554.029 &  0.00 & 0.170   &  $<1.50$ & $< 0.20$ &  & 2.17 \\ 
\hline\noalign{\smallskip}
\multicolumn{7}{c}{Molecular bands}\\
\hline\noalign{\smallskip}
element & \multicolumn{2}{c}{molecule} &\multicolumn{2}{c}{ band } & A(X) &  A(X)\sun\\
\hline\noalign{\smallskip}
C  &\multicolumn{2}{c}{ CH }&\multicolumn{2}{c}{ G-band} & 7.70 & 8.50\\
\hline\noalign{\smallskip}
\multicolumn{6}{l}{$^a$ We list the strongest line of the doublet.}\\
\multicolumn{6}{l}{$^b$ Derived from the EW and the fitting formula}\\
\multicolumn{6}{l}{\phantom{$^b$}of Sbordone et al. (2010).} \\
\multicolumn{7}{l}{$^c$ 1 $\sigma$}\\
\multicolumn{7}{l}{$^d$ 3 $\sigma$}\\
\end{tabular}
\end{table}

\begin{table*}
\caption{\label{leo2} Abundances of SDSS\,J1035+0641}
\centering
\setlength{\tabcolsep}{2pt}
\begin{tabular}{lccrlllll}
\hline\noalign{\smallskip}
\multicolumn{7}{c}{Atomic lines}  \\
\hline\noalign{\smallskip}
Ion & $\lambda$ &$\chi$ & log gf & EW & A(X) & A(X) & A(X)$_{err}$ & A(X)\sun\\
    & nm        & eV    &        & pm &       log g = 4.0     & log g = 4.4 &             &   \\
\hline\noalign{\smallskip}
\ion{Li}{i} & 6707.761$^a$ &  0.00 &--0.009  & $<1.80 $& $<\phantom{-}1.1^b$ &$<\phantom{-}1.1^b$ &  & 1.03 \\
\ion{Mg}{i} & 518.3604 &  2.72 & --0.239 & $<0.30$ & $<\phantom{-}2.47$ & $<\phantom{-}2.47$  &  & 7.54\\
\ion{Ca}{ii}& 393.3663 &  0.00 &  0.105  &  syn  & $\phantom{<-}1.35$ & $\phantom{<-} 1.53$ & 0.10  & 6.33\\
\ion{Fe}{i} & 382.0425 &  0.86 &  0.119  &  $<0.57^c$ & $<\phantom{-}1.91$ & $<\phantom{-}1.91$ &  & 7.52\\ 
\ion{Fe}{i} & 382.0425 &  0.86 &  0.119  &  $<1.67^d$& $<\phantom{-}2.45$ &  $<\phantom{-}2.45$  &  & 7.52\\ 
\ion{Sr}{ii}& 407.7709 &  0.00 &  0.167  &  $<0.50$ & $<-1.80$ & $<-1.67 $ & & 2.92\\
\ion{Ba}{ii}& 4554.029 &  0.00 & 0.170   &  $<3.30$ & $< -0.49$ & $< -0.38$ & & 2.17 \\ 
\hline\noalign{\smallskip}
\multicolumn{7}{c}{Molecular bands}\\
\hline\noalign{\smallskip}
element & \multicolumn{2}{c}{molecule} &\multicolumn{2}{c}{ band } & A(X) & A(X) &  &A(X)\sun\\
        &      & &                     &                         & log = 4.0 & log = 4.4 \\
\hline\noalign{\smallskip}
C  &\multicolumn{2}{c}{ CH }&\multicolumn{2}{c}{ G-band} & 6.90 & 6.70 & & 8.50\\
\hline\noalign{\smallskip}
\multicolumn{6}{l}{$^a$ We list the strongest line of the doublet.}\\
\multicolumn{6}{l}{$^b$ Derived from the EW and the fitting formula}\\
\multicolumn{6}{l}{\phantom{$^b$}of Sbordone et al. (2010).} \\ 
\multicolumn{7}{l}{$^c$ 1 $\sigma$}\\
\multicolumn{7}{l}{$^d$ 3 $\sigma$}\\
\end{tabular}
\end{table*}

% Table 8
\begin{table}
\caption{\label{abuJ1137} Mean abundances of SDSS\,J1137+2553}
\centering
\setlength{\tabcolsep}{2pt}
\begin{tabular}{lccrcc}
\hline\noalign{\smallskip}
Ion                & A(X)   &A(X)\sun& [X/H]    & $\sigma$ & N lines\\
\hline\hline\noalign{\smallskip}
\ion{Li}{i}        & 2.26   & 1.03   &          &      & 1\\
\ion{O}{i}         & 7.18   & 8.76   &   --1.58 &      & 3\\
\ion{Na}{i}        & 6.20   & 6.30   &   --0.10 &      & 2\\ 
\ion{Mg}{i}        & 6.17   & 7.54   &   --1.37 & 0.09 & 3\\
\ion{Al}{i}        & 3.30   & 6.47   &   --3.17 &      & 1\\
\ion{Si}{i}        & 5.10   & 7.52   &   --2.42 &      & 1\\
\ion{Ca}{i}        & 4.20   & 6.33   &   --2.13 &      &2\\
\ion{Ca}{ii} K     & 4.15   & 6.33   &   --2.18 &      &1\\
\ion{Ca}{ii} IR    & 4.60   & 6.33   &   --1.73 &      & 1\\
\ion{Sc}{ii}       & 1.03   & 3.10   &   --2.07 &      & 1\\
\ion{Ti}{ii}       & 2.49   & 4.90   &   --2.41 & 0.21 &11\\
\ion{Cr}{i}        & 3.03   & 5.64   &   --2.61 &      & 1\\
\ion{Fe}{i}        & 4.86   & 7.52   &   --2.66 & 0.17 & 43\\
\ion{Fe}{ii}       & 4.84   & 7.52   &   --2.68 & 0.22 &  6\\
\ion{Co}{i}        & 2.97   & 4.92   &   --1.95 &      & 1\\
\ion{Ni}{i}        & 3.35   & 6.23   &   --2.88 & 0.13 & 2\\
\ion{Sr}{ii}       & 1.50   & 2.92   &   --1.42 & 0.14 & 2\\
\ion{Ba}{ii}       & 2.31   & 2.17   &    +0.14 & 0.32 & 6\\
\hline\noalign{\smallskip}
\multicolumn{6}{c}{Molecular bands}\\
\hline\noalign{\smallskip}
 C --CH (G-band)     & 8.60 & 8.50     & +0.10 & & \\
 N --CN (388nm)      & 7.86 & 6.88     & -0.98 & & \\     
\hline\noalign{\smallskip}
\end{tabular}
\end{table}

% Table 9
\begin{table}
\caption{\label{abuJ1245} Mean abundances of SDSS\,J1245-0738}
\centering
\setlength{\tabcolsep}{2pt}
\begin{tabular}{lccrcc}
\hline\noalign{\smallskip}
Ion                & A(X)  &A(X)\sun& [X/H]     & $\sigma$ & N lines\\
\hline\hline\noalign{\smallskip}
%\ion{Li}{i}        &        & 1.03    &       & & 1\\
\ion{Na}{i}        & 4.30   & 6.30   & --2.00 &      & 2\\ 
\ion{Mg}{i}        & 5.05   & 7.54   & --2.49 &      & 2\\
\ion{Al}{i}        & 3.10   & 6.47   & --3.37 &      & 1\\
\ion{Si}{i}        & 4.30   & 7.52   & --3.22 &      & 1\\
\ion{Ca}{i}        & 3.53   & 6.33   & --2.80 &      & 2\\
\ion{Ca}{ii} H\&K & 3.98   & 6.33   & --2.35 &      & 2\\
\ion{Ca}{ii} IR    & 4.20   & 6.33   & --2.13 &      & 1\\
\ion{Ti}{ii}       & 2.10   & 4.90   & --2.80 & 0.25 & 7\\
\ion{Fe}{i}        & 4.27   & 7.52   & --3.25 & 0.48 &29\\
\ion{Fe}{ii}       & 4.42   & 7.52   & --3.10 & 0.17 & 3\\
\ion{Sr}{ii}       &--0.38  & 2.92   & --3.30 &      & 2\\
\ion{Ba}{ii}       & 0.23   & 2.17   & --1.94 &      & 2\\
\hline\noalign{\smallskip}
\multicolumn{6}{c}{Molecular bands}\\
\hline\noalign{\smallskip}
 C --CH (G-band)     & 8.65 & 8.50     & +0.15 & & \\
\hline\noalign{\smallskip}
\end{tabular}
\end{table}

\subsubsection{SDSS\,J1035+0641}
For this star we have at our disposal {both the  X-Shooter
and UVES
spectra}. The SDSS spectrum of SDSS\,J1035+0641 shows no detectable G-band
and the star was targeted for the weakness of its  \ion{Ca}{ii} K line 
(see Fig.\,\ref{hydra_leo2}), 
yet the X-Shooter
spectrum allows us to detect a definite G-band, that qualifies the star as CEMP
star with A(C)=6.80.
The analysis of the UVES spectrum yields A(C)=6.9, confirming the result
from the X-Shooter spectrum.
The only  metallic features  measurable
on our spectra (both X-Shooter and UVES) are the \ion{Ca}{ii} H\&K lines.
In Table \ref{leo2} we provide the abundances of Ca and C and several
upper limits, both for the assumption log g = 4.0 and log g =4.4.
The upper limits for Li, Mg, Fe, and Sr are derived from the UVES spectrum, 
while the limit on  Ba relies only on the X-shooter spectrum, since
the strongest \ion{Ba}{ii} resonance line is not covered
by the UVES spectrum.
In the discussion we assume the values corresponding to log g = 4.0,
since we consider this option makes the star brighter and its discovery
more likely.
For log g = 4.0 the Ca abundance
A(Ca)=1.35, corresponding to [Ca/H]=--5.0.
In Fig.\,\ref{leo2Fe} we show the observed r spectra in a region
covering three of the strongest \ion{Fe}{i} lines and a synthetic
spectrum corresponding to [Fe/H]=--4.5, none of the three lines
can be confidently detected.
This is one of the five most iron poor stars found so far. 
Observations at higher resolution and S/N ratio are needed to 
determine the iron abundance of this star.
No Li is detected in this star, however our upper limit 
confirms that the star is below the {\em Spite plateau}.
An observation of the Li region with higher  S/N is
highly desirable.

\subsubsection{SDSS\,J1137+2553}

In spite of the 
poor S/N ratio of the only spectrum available for this star, 
the high resolution and wide
spectral coverage of the
UVES spectrum produced an almost complete chemical inventory
summarised in Table \ref{abuJ1137}. 
The C abundance is essentially solar while the oxygen abundance 
is 1.6\, dex below solar and the N abundance only 1\,dex below solar.
This star, with $[{\rm Fe/H}]=-2.70$,
is the most metal-rich star in the present sample.
Sodium is remarkably enhanced with respect to iron, 
with essentially a solar abundance.
 Magnesium also appears to be strongly enhanced, about 1.5\,dex more than iron.
The Sr and Ba abundances measured in this star suggest  that it should be  classified 
as  a CEMP-rs star, using the criterion of \citet{masseron10}. 
Another remarkable feature of this star is its Li abundance, 
with A(Li)=2.26 it lies clearly on the {\em Spite plateau}, unlike 
the majority of CEMP dwarf stars where Li is not measured or appears
clearly below the {\em Spite plateau} \citep{sivarani,behara}.

\subsubsection{SDSS\,J1245-0738}

The quality of the available spectrum is very poor, yet with
the rebinning we were able to detect many metallic lines and confidently
measure the carbon abundance from the G-band.
It was possible to measure the abundance of nine elements, given
in Table \ref{abuJ1245}, and in spite
of the large line-to-line scatter we  derived $[{\rm Fe/H}]=-3.21$ from
thirty \ion{Fe}{i} lines.
The carbon abundance is roughly solar in this star as well, and both Na
and Mg are strongly enhanced with respect to iron.
The Sr and Ba abundances allow us to classify the  star as a
CEMP-s star. 
Our conclusion on the CEMP-s nature of this
star agrees with that of \citet{aoki13}. For all the neutral atoms (Na, Mg, Fe) our results agree
very well with the results of \citet{aoki13}. Their
Ca abundance closely agrees  with the value we determine from the \ion{Ca}{ii} H and K lines;
however, there are strong discrepancies in the abundances of C, Sr, and Ba, and to a
lesser extent of Ti. These  differences are due
to the different gravity values adopted in the two studies, corresponding to the HB status that we prefer 
and the TO status preferred by \citet{aoki13}; the ionised atomic species are
sensitive to the adopted gravity and so is the G-band.

\subsubsection{SDSS\,J1742+2531 \label{ercolinachem}}
In spite of the higher resolution of our UVES spectrum of SDSS\,J1742+2531
compared to the X-Shooter spectrum presented in 
\citet{gto13}, we can confidently detect only
five metallic lines in addition to \ion{Ca}{ii} H and K, which were also detected in the X-Shooter 
spectrum:  the
two strongest lines of the infrared \ion{Ca}{ii}
triplet
and three \ion{Fe}{i} lines. 
These are listed in Table \ref{abund}.
We provide the abundances implied by our spectra
both for log g = 4.0 and log g = 4.3.

The Ca abundance is very coherent among 
\ion{Ca}{ii} H and K and the \ion{Ca}{ii} IR triplet.
The mean Ca abundance
of all  four of the available lines is $\langle {\rm A(Ca)} \rangle = 1.77$
with a line-to-line standard deviation of 0.03\,dex.
This corresponds to 
[Ca/H] = --4.56 and
is consistent with the Ca abundance that we
derived only from the \ion{Ca}{ii} H and K in
the X-Shooter spectrum ([Ca/H]$\le -4.5$; \citealt{gto13}). 
The \ion{Ca}{i} 422.6\,nm line is not detected. 
The departures from  LTE of the Ca lines in metal-poor
stars have been extensively discussed in the
recent literature \citep{spiteCa,Korn09,Mash07}
and it is recognised that the effects on the \ion{Ca}{ii} H and K lines is very small,
while it can be significant for the  \ion{Ca}{ii} IR triplet.

The gravity sensitivity  of  the \ion{Ca}{ii} H and K lines
and of the   \ion{Ca}{ii} IR triplet
is very different, the first two being essentially
insensitive to gravity.
We  checked the effect of a gravity as low as log g = 3.3
on the Ca abundances, the mean Ca abundance derived
from the  \ion{Ca}{ii} H and K lines
is lower by 0.04\,dex, while 
that deduced from the 
\ion{Ca}{ii} IR triplet
is 0.22\,dex lower.
This gives a discrepancy of the order of 0.2\,dex
between the two transitions. Although this would still be within the
errors-- the line-to-line scatter would rise to only 0.1\, dex and the
calcium abundance would fall  to 1.63-- we believe it points
towards a turn-off gravity around log g = 4.0.

The iron abundance 
derived from the \ion{Fe}{i} lines is also very consistent
among the three lines 
and is $[{\rm Fe/H}]=-4.80$ with a standard deviation of 0.07\,dex.

The analysis of the G-band confirms the carbon abundance
we derived from the X-Shooter spectrum (A(C)=7.4 \citealt{gto13})
within errors  A(C)=7.26$\pm 0.2$. 
The quality of the spectra is not high enough to
place any constraint on the $^{12}$C/$^{13}$C ratio.
We inspected our spectra to see if we could place
meaningful upper limits on other elements and we provide  
these in Table \ref{abund}, the limit
on the  EWs corresponds to a 3$\sigma$ detection and
$\sigma$ has been estimated from the Cayrel
formula \citep{cayrel88}.
We were not able to place a meaningful limit on the
nitrogen abundance either from the UV NH band
at 336\,nm
or from the violet CN band at 388\,nm, as these would require
a much higher S/N than provided by our spectra.

The \ion{Li}{i} resonance doublet at 670.7\,nm
is not detected; the Cayrel formula and the measured 
S/N ratio imply that its EW is less than 0.83\,pm (3$\sigma$ criterion). The upper limit
on the Li abundance was derived
from the fitting formula of
\citet{sbordone10}  
providing the  3D NLTE Li abundance. The 1D LTE or NLTE estimate
would differ by only 0.03 dex, thus it is irrelevant in 
the present context.

\begin{table*}
\caption{\label{abund} Abundances of SDSS\,J1742+2531 }
\centering
\setlength{\tabcolsep}{2pt}
\begin{tabular}{lccrrrrrrc}
\hline\noalign{\smallskip}
\multicolumn{7}{c}{Atomic lines}\\
\hline\noalign{\smallskip}
Ion & $\lambda$ &$\chi$ & log gf & EW & EW$_{\rm err}$& A(X) & A(X) & A(X)$_{err}$ & A(X)\sun\\
    & nm        & eV    &        & pm &  pm           & log g = 4.0 & log g = 4.3   &   &   \\
\hline\noalign{\smallskip}
\ion{Li}{i} & 6707.761$^a$ &  0.00 &--0.009  & $<0.83 $& & $<1.8^b$  & $< 1.8^b$ &  & 1.03 \\
\ion{O}{i}  & 777.1941 &  9.15 & 0.369   & $<1.00$ & & $<6.92$ & $<6.97$ &  & 8.76\\
\ion{Na}{i} & 588.9951 &  0.00 & 0.117   & $< 0.90$& & $< 2.14$& $<2.14$ &  &6.30\\  
\ion{Mg}{i} & 518.3604 &  2.72 & --0.239 & $<1.00$ & & $<3.07$ & $<3.07$ &  &7.54\\
\ion{Si}{i} & 390.5523 &  1.91 & --1.041 &  syn    & & $<3.05$ & $<3.05$ &  & 7.42\\
\ion{S}{i}  & 921.2863 &  6.53 & 0.420   & $<1.10$ & & $<4.61$ & $<4.75$ &  &7.16\\
\ion{Ca}{i} & 422.6728 &  0.00 &  0.265  & $<1.00$ & & $<1.62$ & $<1.62$ &  &6.33\\
\ion{Ca}{ii}& 393.3663 &  0.00 &  0.105  &  syn  &  &1.79 & 1.81 & 0.04 &6.33\\
\ion{Ca}{ii}& 396.8469 &  0.00 & --0.200 &  syn  &  &1.76 & 1.79 & 0.08 &6.33\\
\ion{Ca}{ii}& 854.2091 &  1.70 & --0.514 &  6.20 & 0.29 &1.72 & 1.81  & 0.05 &6.33\\
\ion{Ca}{ii}& 866.2141 &  1.69 & --0.770 &  5.20 & 0.29 &1.79 & 1.88  & 0.05 &6.33\\
\ion{Fe}{i} & 382.0425 &  0.86 &  0.119  &  2.40 & 0.35 & 2.73 & 2.73 & 0.09 &7.52\\ 
\ion{Fe}{i} & 382.5881 &  0.92 & --0.037 &  1.90 & 0.35 & 2.80 & 2.80 & 0.09 &7.52\\ 
\ion{Fe}{i} & 385.9911 &  0.00 & --0.710 &  1.90 & 0.35 & 2.63 & 2.63 & 0.09 &7.52 \\
\ion{Sr}{ii}& 407.7709 &  0.00 &  0.167  &  $<1.40$ &  & $<-1.25$ &$<-1.16$ & 2.92\\
\ion{Ba}{ii}& 4554.029 &  0.00 & 0.170   &  $<1.50$ &  & $<-0.97$ &$<-0.84$ & 2.17 \\ 
\hline\noalign{\smallskip}
\multicolumn{7}{c}{Molecular bands}\\
\hline\noalign{\smallskip}
element & \multicolumn{2}{c}{molecule} &\multicolumn{2}{c}{ band } & A(X) & A(X) & &  A(X)\sun\\
        &   & &                        & & log = 4.0 & log g = 4.3 \\
\hline\noalign{\smallskip}
C  &\multicolumn{2}{c}{ CH }&\multicolumn{2}{c}{ G-band} & 7.26 & 7.08 & & 8.50\\
\hline\noalign{\smallskip}
\multicolumn{6}{l}{$^a$ We list the strongest line of the doublet.}\\
\multicolumn{6}{l}{$^b$ Derived from the EW and the fitting formula}\\
\multicolumn{6}{l}{\phantom{$^b$}of Sbordone et al. (2010).} 
\end{tabular}
\end{table*}

\subsection{Error estimates}

The observational material presented here is heterogeneous
as far as resolving power and signal-to-noise ratio are concerned.
It is therefore not simple to provide a consistent
estimate of the errors for our abundances.
For the stars for which several lines of a given element are
measured we provide the standard deviation of the measurements
and that can be taken as an estimate of the statistical error on the
abundance. When several lines are measured and no $\sigma$ is 
provided (e.g. the \ion{O}{i} triplet or \ion{Ca}{ii}  H\&K
lines), this means that the lines have been fitted together.
In these cases we suggest taking the line-to-line scatter
of the \ion{Fe}{i} lines as an estimate of the error because  the statistical error is linked
to the signal-to-noise ratio of the spectra, \ion{Fe}{i} lines
are found over a large wavelength range, and we assume that
our measured line-to-line scatter is dominated by the noise
in the spectrum.

The situation is less clear for the stars for which only a few lines
can be measured. For these stars for each measured line we provide an estimate of
the error on the equivalent width derived from
the Cayrel formula \citep{cayrel88} and the measured signal-to-noise ratio
in the spectrum. From this estimate of the error on the
equivalent width we also derive an estimate of the error
on the abundance. For the \ion{Ca}{ii} H and K lines
for which the abundance has been derived using
line profile fitting, we estimate the error
as the corresponding diagonal element in the
covariance matrix of the fit, ignoring all the off-diagonal terms.

For the carbon abundances (and nitrogen abundances from the CN band), 
the error is dominated by the placing of the continuum.
Our fitting was limited to the blue part of the band, excluding the
band-head. Given that the fit covers a large wavelength range,
the error in the continuum placement is not  dominated by the signal-to-noise
ratio (the noise is averaged out over so many pixels, yet it does play a role), but rather by
the quality of spectrum rectification (removal of the blaze function).
In \citet{Spite06} error estimates on the fitting of
the G-band are provided, resulting from fits derived by fixing
slightly different continuum levels. In that case the errors
are of the order of 0.1\,dex.
In the present case the data are of much lower S/N ratio, and for
the X-Shooter spectra  resolving power as well. 
We estimate 0.15\,dex, which is a conservative estimate of the error
on our C abundances.

The systematic errors, i.e. the effect of the adopted
atmospheric parameters on the derived abundances have  often
been discussed in the literature; we point to the reader Table 4 of
\citet{bonifacio09} where this exercise is done for a star
with atmospheric parameters close to the stars discussed
in the present paper. The systematic errors in our case are of the
same order of magnitude.
In that table the systematic error on the C abundances derived from
the G-band is not discussed: a change in temperature of 100 K
corresponds to a change of 0.17\,dex in C abundance.

\subsection{Effects of departures from local thermodynamic equilibrium}

The deviations from LTE in
the formation of \ion{Ca}{i} and \ion{Ca}{ii} lines
in metal-poor stars has been extensively studied by
\citet{Mash07} and  \citet{spiteCa}.
It has been recognised that the \ion{Ca}{i} 422.6\, nm
resonance line provides discrepant results even when computed
in NLTE \citep{spiteCa}, for this reason we never used the abundance from this
line even when it was measured.
On the other hand, both \ion{Ca}{ii} K line and and the IR triplet 
lines can provide reliable Ca abundances. While the K line
is always formed quite close to LTE conditions, the IR triplet
shows significant deviations from LTE. The NLTE corrections
are always negative.
This can be appreciated in the three stars
SDSS\,J0212+0137, SDSS J1137+2553, and SDSS J1245-0738. 
In all three stars the IR triplet provides Ca abundances
that are higher by 0.2 to 0.4\, dex than the K line. 
This is compatible with expected NLTE corrections.
According to the 
computations of \citet{Mash07}, at a metallicity of --3.0 for \teff = 6000 and log g = 4.0
the non-LTE correction for the IR triplet is --0.23\,dex. 
Star CS 29527-015, analysed by \citet{spiteCa},
has parameters very close to those of 
SDSS\,J0212+0137 and the computed NLTE correction is  --0.51\,dex.

The exception is SDSS J1742+2531, for which the LTE analysis
provides the same Ca abundance  from the H and K lines 
and from the IR triplet. Although the star has almost the same
atmospheric parameters as the three above-mentioned stars,
we note that 
its Ca abundance is much lower, thus the equivalent width
of the IR triplet lines is smaller. Generally speaking,  as the equivalent width of a line decreases
it forms closer to LTE. Although we do not have in hand
specific computations for the \ion{Ca}{ii} IR triplet
to demonstrate this effect, we know it can be seen for other lines (see e.g. Fig.  6 in \citet{AndrievskyNa} illustrating
this behaviour for the \ion{Na}{i} D lines).
It is thus quite likely that the NLTE corrections
for the IR Ca triplet in  SDSS J1742+2531 are much smaller
than for the other three stars

Regarding NLTE effects on neutral iron 
for star HE\,1327-2326, \citet{aoki} 
adopted a correction of +0.2\,dex, based on several previous
investigations \citep{gratton99,Korn03,theve}.
This value should also be appropriate  
for SDSS\,J1742+2531, which has similar atmospheric parameters.

\subsection{Effects of granulation}  

It  is well known that 
the formation of molecular bands in the atmospheres of metal--poor
stars is prone to strong granulation effects \citep[see e.g.][and references therein]{bonifacio13}.
From  Table 6 in \citet{bonifacio09} we estimate that a correction between --0.5 and --0.6\, dex
in the carbon abundance derived from the G-band is necessary
for all the stars here investigated.

For the three lines of \ion{Fe}{i} measured in SDSS J1742+2531, the 3D correction is about --0.4\,dex,
consistent with the    \citet{bonifacio09} results.
This implies that NLTE and granulation effects 
have opposite directions, 
and they may tend to cancel; however, for iron it is not
possible to simply add the 3D and NLTE corrections \citep{Mash13}.
The granulation effect for \ion{Ca}{ii} lines is small, of the order of --0.1\,dex \citep{Caffau12}.

\section{On the carbon abundances in CEMP stars}
\label{cplat}

Of the six stars analysed in the present paper, two --
SDSS\,J1137+2553 and SDSS\,J1245-0738 -- have been classified
as CEMP-rs and CEMP-s, respectively. 
For the purposes of this discussion we  consider
these objects as a single class (CEMP-rs+CEMP-s), subdivided into two
different
subclasses (CEMP-s and CEMP-rs). 
As detailed below, 
there are reasons to believe 
these stars are the result of mass-transfer
from an AGB companion in a binary system. For these two
particular stars, this may be at odds with the lack of obvious
radial velocity variations, yet  few measurements are
available.  One cannot exclude the hypothesis that both
systems are seen nearly face-on, hence with little or no radial velocity variations.

We now concentrate our discussion on the four other stars, of which 
SDSS\,J0212+0137 can be classified as a CEMP-no star.
For the three other stars it is not possible to establish
whether they are CEMP-s or CEMP-no on the basis of the
abundances  of  Ba because the available upper limits
do not allow us to distinguish between the two classes.
However, as discussed in Sect.\,\ref{chigh},  
on the basis of their carbon
abundance we have reasons to believe that they are indeed
CEMP-no stars.

%%% FIGURE %%%%%%%%%%%%%%%%
\begin{figure}
\begin{center}
\resizebox{\hsize}{!}{\includegraphics[clip=true]{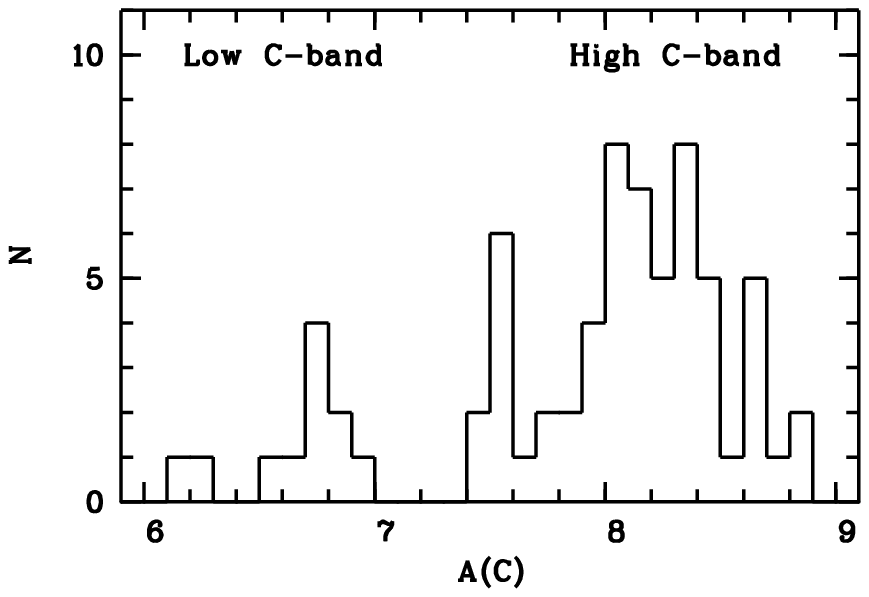}}
\end{center}
\caption{Carbon abundances in all the CEMP stars 
in the present paper and in the literature
with $[{\rm Fe/H}]>-3.5$ 
\citep{sivarani,thompson08,aoki08,behara,placco,carollo,masseron10,masseron12,yong2013,cohen2013,Spite13,gto13}.
The bimodal distribution is clear in this plot.
We refer to the peak at lower carbon abundance as the {\em low-carbon band}
and to the peak at higher carbon abundance as the {\em high-carbon band}.
}
\label{plotc_histo}
\end{figure}
%%% FIGURE %%%%%%%%%%%%%%%%

An interesting feature  displayed by the
CEMP stars is shown in Fig. \ref{plotc_histo}. 
There are many known CEMP stars with $[{\rm Fe/H}]>-3.5$.
If we plot the histogram of the carbon abundances
in these stars, we see that the distribution
is clearly bimodal (Fig.\,\ref{plotc_histo}). 
Each of the two peaks is quite wide, almost
one dex, but they appear to be rather well separated. 
In the following we shall refer to these two peaks in carbon
abundance 
as the {\em high-carbon band} and the {\em low-carbon band}.
Quite interestingly all stars with $[{\rm Fe/H}]<-3.5$ belong
to the  {\em low-carbon band}.
Another way to look at the carbon abundances is to plot
the carbon abundance versus [Fe/H], as in Fig.\,\ref{plotc}.
Similar
plots have already been shown in \citet{behara}, \citet{masseron10}, \citet{Spite13} and
\citet{gto13}.

In Fig.\,\ref{plotc} we have included only unevolved 
stars since giant stars may be internally
``mixed'' \citep{Spite05,Spite06}  and the
C abundance may thus be  decreased by an amount 
that is generally difficult to estimate.
We have included in the plot only stars from the literature that have a measured
Ba abundance or a significant upper limit
\citep{sivarani,frebel05,frebel06,thompson08,aoki08,behara,masseron10,masseron12,yong2013,cohen2013,Li2015}, 
so as to have some
indication on their classification as CEMP-no, CEMP-s, or CEMP-rs.
In order to clearly define the behaviour of the carbon abundance at low
metallicity, 
we have also included  the ultra-iron-poor  subgiant
HE 1327-2326 \citep{frebel05,frebel06,cohen2013} and
SDSS\,J1619+1705 \citep{gto13}, even though their Ba abundance is undetermined.
We have also included  the lower RGB giants 
SMSS J0313-6708 \citep{Keller},  
HE\,0107-5240 \citep{chris2004}  and HE\,0557-4840 \citep{norris07},
which are in the CEMP-no class. 
These 
stars are not more luminous than the RGB ``bump'' and therefore
should not be deeply ``mixed''; however, they have
undergone the first dredge-up and following \citet{bonifacio09}
their carbon abundance should have been lowered by 0.26 dex
compared to any TO star.

The new determination of A(C)
for SDSS\,J1742+2531, based on the UVES spectra,
has replaced the earlier value based on X-Shooter.
These new observations, especially that of
 SMSS J0313-6708, confirm that all
the lowest metallicity stars belong to the  
{\em low-carbon  band}.
There is a wide scatter (over 1\,dex) in carbon abundances 
among these stars, yet there is no clear trend with
[Fe/H] and they are all well below the {\em high-carbon band}
defined by the other CEMP stars, which is almost
at solar C abundance.
Here we  redefine the class of CEMP-no stars
as $\rm [Ba/Fe] \leq 1.2$.
The limit of [Ba/Fe] = 1.2 was chosen because it is the highest value reached (LTE computation)
by the normal metal-poor stars (not C-rich) 
\citep{Hill02,F2007,Spite14}.
This new definition also  has  the virtue of allowing us
to classify  SDSS\,J1036+1212
and CS\,29528-41 as CEMP-no; they  had previously been classified as CEMP-no/s and CEMP-s by
\citet{behara} and \citet{sivarani}, respectively.  
The new definition does not require   a new
class to be created for these two objects.

The vast majority (91\%) of stars on the
{\em high-carbon band} are  of the type CEMP-s or CEMP-rs; 
this population of stars is compatible with a 
population of 100\% binaries of shorter periods,
at a maximum of 20\,000 days (\citealt{Starkenburg},
see also \citealt{lucatello}).
The stars of the {\em low-carbon band} for which Sr and Ba
have been measured are all of the CEMP-no type.
Although two of them have been found to be binaries
\citep{Starkenburg}, the binary frequency is similar
to that  observed in the solar neighbourhood.
A very clear picture is emerging from the observations and  we 
propose an interpretation.

\subsection{High-carbon band} \label{chigh}

%%% FIGURE %%%%%%%%%%%%%%%%
\begin{figure*}
\begin{center}
\resizebox{\hsize}{!}{\includegraphics[clip=true]{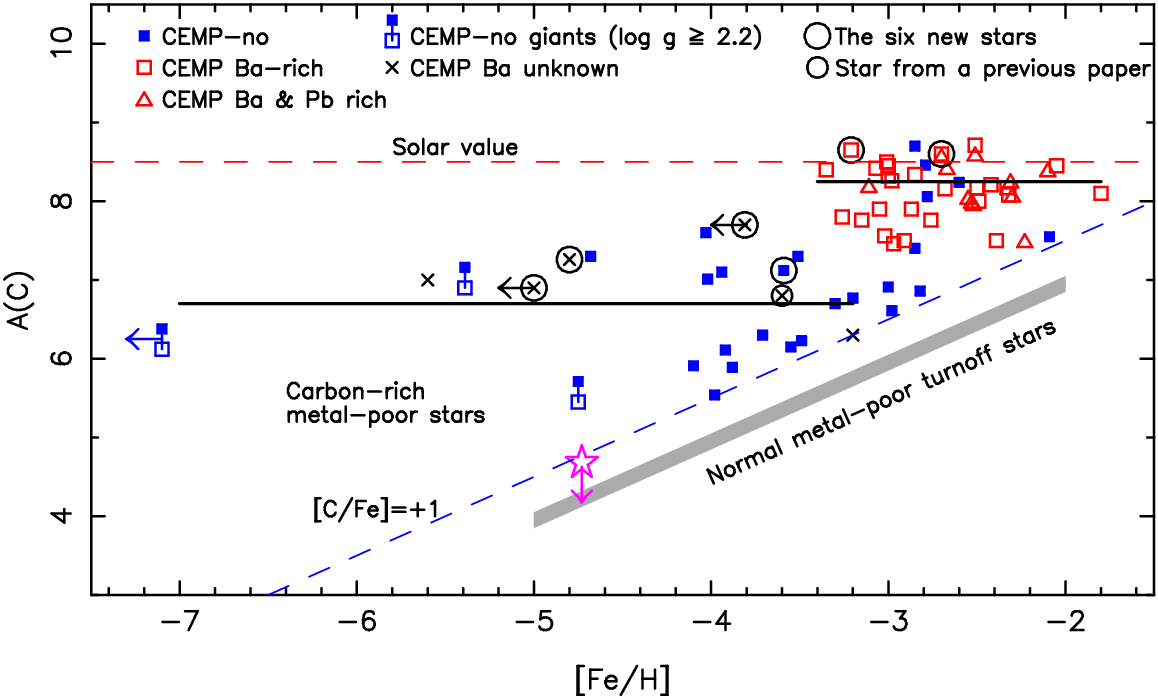}}
\end{center}
\caption{ The carbon abundances A(C) of CEMP
stars as a function of [Fe/H]. 
The stars  in the present paper 
and in \citet{gto13}
are shown with big   and
small circles, respectively.
The other turn-off stars come from the literature
\citep{sivarani,frebel05,frebel06,thompson08,aoki08,behara,masseron10,masseron12,yong2013,cohen2013,Li2015}. 
The CEMP-no turn-off stars are
represented by  filled blue squares.
The CEMP-no giants are
represented by two symbols: an open blue square for the measured value
of the C abundance, and a filled blue square for the empirically corrected (for the first dredge-up) A(C) value.
The CEMP-s or -rs stars  are indistinctly represented by 
 open red squares, but by  open red triangles if they are also Pb-rich.
When the Ba abundance is unknown the star is marked with a black 
cross.
The upper limit marked as a violet star is the only
``C-normal'' star that appears in this plot: SDSS\,J1029+1729 \citep{Caffau12}.
The
solid horizontal (black) lines 
represent the mean value of the carbon abundance for CEMP stars with $\rm [Fe/H] > -3.0$ (A(C)=8.25) and 
for CEMP stars with $\rm [Fe/H] < -3.4$ (A(C)=6.8) as derived by
\citet{Spite13}.
On the upper A(C) band all the stars but four (HE\,0007-1832, HE\,022-4831,
SDSS\,J0036-1043, and CS\,22958-42) are Ba-rich.  On the contrary, on the
lower A(C) band all the stars are CEMP-no.
}
\label{plotc}
\end{figure*}
%%% FIGURE %%%%%%%%%%%%%%%%

As a consequence
of the results of \citet{Starkenburg}
and \citet{lucatello} on the binarity
of the CEMP-s stars (100\% of binaries), 
we assume that
the {\em high-carbon band} CEMP stars
are all post-mass transfer  binary systems.
The originally more massive star of the system
has gone through the AGB phase and 
 donated mass to its companion, and subsequently evolved
to the present white dwarf status.
The high-carbon abundances in these 
stars have three causes:
{\em i)} the AGB stars enrich their atmospheres
through the third dredge-up of carbon produced in the
He-burning shell, the amount of carbon produced
and dredged up is independent of the star's initial
metallicity;
{\em ii)} since the binary system is extremely metal-poor,
once this material is transferred to the companion, the transferred
carbon exceeds 
the carbon originally present in the companion's atmosphere by at least an order of magnitude (see Fig.\,\ref{plotc}),
thus  the metallicity of the system is again irrelevant with 
respect to the final carbon abundance in the companion;
 and {\em iii)} the dilution factor\footnote{Defined as  the base 10 logarithm of the mass of the convective
envelope of the star to the mass of material accreted by the AGB companion}
varies over a limited range.

This last requirement  is supported
by the study of
\citet{bisterzo}, who have analysed  a sample of
100 CEMP-s stars for which detailed chemical
abundances are available (almost all the stars
that appear in Fig.\,\ref{plotc} are in their sample) 
and showed that
the abundances of light elements (C,N,O, Na,  Mg) and neutron capture elements
can be reproduced assuming transfer from an AGB star
in the mass range 1.3-2\,M$_\odot$\ and 
logarithmic dilution factors in the range 0.0-2.5.
This is consistent with the constraints deriving from the carbon abundance alone.
The {\em high-carbon band} spans the range A(C)=7.4--8.9, the same range covered by 
the carbon abundances in the AGB donor, according to Table 6 of \citet{bisterzo10}.
The range of masses of the primary star is 
narrowed by the nucleosynthesis
requirements;  the mass of the 
secondary is even narrower. Observations indicate that they are all 
of spectral type F or G, which means, given the old age presumed
from the low metallicity, they are in the mass range 0.7M$_\odot$ to 0.9 M$_\odot$.
The study of \citet{masseron10} also supports this point of view; 
 they detect a strong correlation between Ba and C abundances in
CEMP-s stars. They claim that this supports
the operation of a $^{13}$C neutron source in the AGB companions that are responsible
for the C and $s-$process enrichment in the CEMP-s stars.
The  orbital parameters must also lie in a relatively
narrow range to allow mass transfer to take place
during the AGB phase of the primary.
The result of \citet{Starkenburg} that
the periods are all shorter than 20\,000 days corroborates
this hypothesis.
We note that  the study of \citet{aoki15} also suggests that the 
binary stars among extremely metal-poor stars are biased towards 
shorter periods.

This does not imply that binary systems are not
created with a wide range of mass ratios and orbital
parameters, but that only a subset of these systems
may evolve in such a way as to give rise to a 
{\em high-carbon band} CEMP star. 
This follows from the results of \citet{lucatello} and \citet{Starkenburg}
on binarity and of \citet{bisterzo} on nucleosynthesis.

This scenario also explains why such stars are not found
at higher metallicities. In that case the
carbon initially present in the secondary's atmosphere
is non-negligible with respect to the amount transferred.
As a consequence the star is not classified as carbon enhanced
because the C/Fe ratio is not very high.
It has already been 
pointed out that these CEMP-s systems are indeed low-metallicity analogues
of CH stars and Ba stars \citep[see][and references therein]{Starkenburg}.
 
Our proposed scenario has interesting implications in the light of the results 
 of some of the current simulations of Pop III star formation 
\citep{clark11,smith11,greif12,dopcke13,stacy13,stacy14}.
The  above  studies unanimously predict a very high binary fraction for Pop III stars and a wide mass 
range with some Pop III stars eventually undergoing the AGB stage.
If such a star is formed in a binary system, it may transfer the metals produced in the AGB
phase to the companion. None of the stars in the {\em high-carbon band}
can be a Pop  III star because there is no way an AGB star can enrich
the companion in Fe. However, this is a motivation for searching for
{\em high-carbon band} stars even at  lower metallicity--to find some Fe-free stars.
According to our proposed scenario 
SMSS J0313-6708 is not one of the Pop III stars that
have accreted mass from an AGB companion because
its C content is too low, unless extensive internal mixing
has decreased the C abundance by about 1\,dex to form N.
This is, however, unlikely considering the surface gravity of
the star (log g = 2.3).

\subsection{Low-carbon band \label{lcband}}

We propose that these stars are not the result of transfer from 
a companion. Some of them may well be members of a binary system
\citep{Starkenburg}, but
no mass transfer has taken place.
The atmospheric abundances of these stars
are {\em \emph{bona fide}} fossil records of the
interstellar medium out of which they were formed.
The variable carbon abundance displayed in these stars is the
result of the nucleosynthesis of a few core collapse SN of zero metallicity that
have polluted the gas. 
The variation in the carbon abundances
reflects a range of masses in the SN progenitors,  as well as
varying degrees of dilution of the SN ejecta with primordial
gas. The spread in Fe abundance reflects a spread in the amount
of fall-back of the SN that produced the bulk of carbon.

None of these stars has yet  been observed
with a C abundance as high as the {\em high-carbon band}, which
places constraints on the minimum dilution possible and
the maximum C production by the SN that polluted the ISM.
Only the product of these two quantities is constrained, however\footnote{It is straightforward
to pass from abundance in number in the usual spectroscopic notation A(X) to the mass
concentration usually used by chemists: M(X)/M(H) $= 10^{(\rm A(X)-12)}\times {m_X\over m_H}$, where
$m_X$ and $m_H$ are the atomic weights of element X and hydrogen, respectively. An abundance
A(C)=6.80 corresponds to a mass concentration of $\sim 7.5\times 10^{-5}$. 
This means that, for example, if a SN ejects a mass of 1 M$_\odot$ of carbon, this has to be
diluted with 1.33$\times 10^4$ M$_\odot$ of hydrogen in order to reach this mass concentration,
or, equivalently, carbon abundance.}.

\subsection{Carbon enhanced damped Ly$\alpha$ galaxies}

We  can use  damped Ly$\alpha$ galaxies (DLAs) 
to determine detailed abundances
at high redshift, and compare the results to the metal-poor
stars in our Galaxy. \citet{Kulkarni} presented a model of chemical
enrichment of galaxies, according to which high
redshift galaxies 
with low-mass haloes (M $\la 10^9$ M$_\odot$)
are sensitive to the  initial mass function (IMF) of Pop III stars.
They tentatively identify such galaxies with DLAs.
No absorption system as metal-poor
as the most metal-poor Galactic stars has yet been found, the record
holder being the Lyman limit system at $z=3.410883$ towards
QSO\,J1134+5742, for which no metallic lines could be detected;
 an upper limit on  the metallicity of $Z< 10^{-4.2}Z_\sun$ has been
determined   \citep{fumagalli}. However,  there is a handful of DLA galaxies 
at $[{\rm Fe/H}]\sim -3.0$ and for a few  C abundances have also been 
measured.
In particular, the $z=2.30400972$ DLA towards  QSO\,J0035-0918
with $[{\rm Fe/H}]=-3.09$ (on our adopted solar abundance scale) has been
claimed to be a CEMP system, in the sense that $[{\rm C/Fe}]>+1$.
The absolute C abundance measured by 
\citet{cooke11} is A(C)=6.92, corresponding to the
previously defined {\em low-carbon band}. The measurement is very delicate because 
 the \ion{C}{ii} lines are saturated and therefore the derived column density
is very sensitive to the adopted turbulent velocity $b$. 
\citet{carswell} revised the measurement, adopting a purely thermal
model and derived A(C)=5.92. \citet{dutta} also analysed this system,
and their preferred model has a very low turbulence ($b=0.89$ \kms)
implying A(C)=5.90, but by adopting $b=2.0$\,\kms\ this value rises to
A(C)=7.63. If the lower C abundances are adopted then the system
 no longer has $[{\rm C/Fe}]>+1$.
What is important for the present discussion is that all the
proposed C abundances are essentially compatible with the 
{\em low-carbon band}, although the lower values are on the low side.
We suggest that this is consistent with our scenario; this DLA galaxy 
should have been enriched by only a few zero-metallicity SNe, at least one of which
must have been faint (see Sect. \ref{cc}). Our scenario would be challenged if the
carbon abundance had been found  on the {\em high-carbon band}, since it 
is unlikely that  a whole galaxy could be polluted essentially by AGB stars.
There is another interesting system, the DLA at $z=3.067295$ towards
QSO\,J1358+6522
\citep{cooke12} that has A(C)=6.18 and $[{\rm Fe/H}]=-2.89$. Again this system
is compatible with the {\em low-carbon band}.
None of the known metal-poor  DLAs shows a carbon 
abundance compatible with the {\em high-carbon band}.
There are carbon-normal DLAs, that fall in the shadowed region
in Fig. \ref{plotc}.
It would be extremely important to detect other DLAs with metallicity as low
as that observed in Galactic stars in order to compare the same 
metallicity regime. 

There is another thing that is intriguing; the metal-poor halo
stars for which ages can be measured always have  ages in excess of 12 Gyr
or even of 13 Gyr, which corresponds to a redshift in excess
of 10. A redshift $z=3$ corresponds to a look-back time of 11.5 Gyr. If our
proposed scenario is correct, the  two  DLAs above have been forming zero-metallicity
stars at very late times. This suggests 
that primordial gas clouds could exist and trigger
star formation down to at least $z=2.3$.   
A similar conclusion has been reached in a completely independent
way by considering the low N/O ratios found in DLAs \citep{molaro2004,zafar}.
We cannot exclude that the stars that enriched  these DLAs  actually formed 
13 Gyr ago, and that the DLA galaxy has not undergone any star formation since.
However, this seems very unlikely. The DLA systems have a high
H column density (hence the damping wings on the  Ly\,$\alpha$ line), comparable to that
of the Milky Way disc if it were    seen face-on.  It is natural
to assume that this also implies  a high volume density, which should
ensure  ongoing star formation as  in the Milky Way disc.
If this is the case the SNe that have enriched the gas we are observing
exploded less than 10$^7$ years earlier.

One cannot exclude that any given DLA is a long filament of low
volume density seen along its length and that it hosts no star formation.
This seems to be a very contrived hypothesis and may only apply 
to a small fraction of the observed DLAs given the difference in
cross section of a gas filament and a star forming galaxy like the Milky Way
or the Magellanic clouds. 
It would be very important to be able to find and study DLAs at higher $z$,
which may be possible thanks to the IR capabilities of the next generation
of 30\,m class telescopes.

The high redshift DLAs at low metallicities 
can also provide us  with some interesting insights into the presence
of dust. We have been invoking dust as the main cooling agent and
DLAs provide evidence for the presence of dust from the abundance ratios
of volatile to refractory elements of similar nucleosynthetic origin,
such as Zn/Fe or S/Si. The refractory elements are usually found to have
lower abundances; however, \citet{molaro06} 
noted from the Zn/Fe ratios  that below $[{\rm Fe/H}]=-2$ there seems
to be no evidence of dust depletion, suggesting a lack of dust
in the lowest metallicity DLAs. A similar result has been found
by \citet{rafelski} for the Cr/Zn ratios.
It is  interesting to note that in 
the $z=3.067295$ DLA towards
QSO\,J1358+6522 the ratio Si/S is essentially solar [Si/S] = 0.07$\pm 0.09$
\citep{cooke12}. 
In the DLA
at  $z=2.30400972$ towards  QSO\,J0035-0918,
the ratio Si/O is again solar, [Si/O]=$+0.06\pm 0.14$ \citep{dutta}.
Although these systems have a low dust content,
it may still be sufficient
to act as a coolant
for the formation of low-mass stars.
 The existing studies of this topic 
\citep{schneider06,omukai08,dopcke11,schneider12,dopcke13}
find that if one assumes that the dust-to-gas ratio D scales with the
metallicity Z, one only needs a metallicity of the order of 
$10^{-5}$ solar to get effective dust cooling. This means
that if we relax the assumption that D scales with Z, 
for example assuming a constant dust-to-gas ratio,
we only need an absolute dust abundance of
the order of $10^{-7}$ solar. 
In gas with gas-phase metallicity  $10^{-3}$ solar, our required dust-to-metals ratio is therefore
only $10^{-2}$, and the associated elemental depletion would not be detectable 
given the current errors on the abundances. 
If we account for the fact that dust grains can grow during the gravitational collapse
of the gas \citep[see e.g.][]{chiaki}, then our required initial dust abundance is
even smaller and this argument is stronger.

%%% FIGURE %%%%%%%%%%%%%%%%
\begin{figure}
\begin{center}
\resizebox{\hsize}{!}{\includegraphics[clip=true]{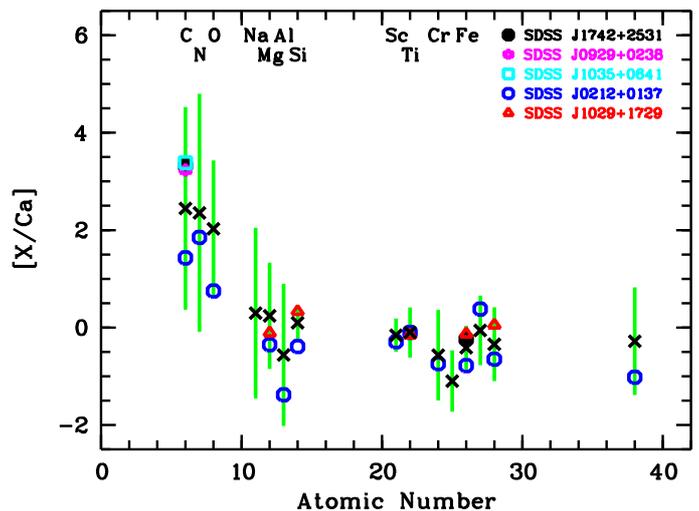}}
\end{center}
\caption[]{The ratios of [X/Ca] for elements carbon to yttrium 
for
SDSS\,J0212+0137 (blue open circle),
SDSS\,J0929+0238 (purple filled circle),
SDSS\,J1035+0641 (cyan open square) and
SDSS\,J1742+2531 (black hexagon).  
These are compared to the most
metal-poor, carbon-normal star SDSS\,J1029+1729 \citep{Caffau12} (red triangles) and to
mean [X/Ca] ratios (black $\times$) in a sample of nine 
extremely metal-poor CEMP stars:
HE\,0107-5240 \citep{chris2004};
HE\,1327-2326 \citep{Frebel08}; HE\,0557-4840 \citep{norris07};
HE\,0134-1519, HE\,0233-0343, HE\,1310-0536 \citep{terese};
 SMSS J0313-6708 \citep{Keller};  G 77-61   \citet{PlezCohen,PCM2005} and  CS 22949-037
\citep{depagne02,norris02}. The last star was excluded
in the computation of the mean and $\sigma$ of the elements C and N 
since it is clearly a ``mixed'' giant, in the sense specified by \citet{Spite05}.
The dispersion around the mean is represented by the 
bars around the $\times$, which correspond to
$2 \sigma$.
For all stars we took the abundances of Ca derived
from the \ion{Ca}{ii} lines, except  for CS 22949-037
for which we took the \ion{Ca}{i}.
}
\label{plotca}
\end{figure}
%%% FIGURE %%%%%%%%%%%%%%%%

%%% FIGURE %%%%%%%%%%%%%%%%
\begin{figure}
\begin{center}
\resizebox{\hsize}{!}{\includegraphics[clip=true]{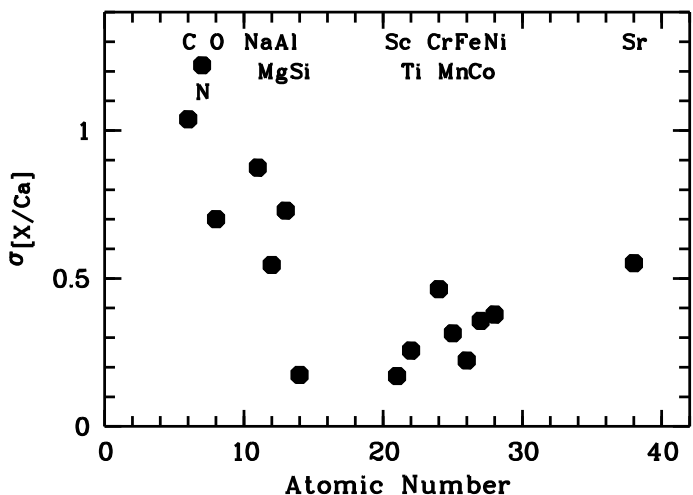}}
\end{center}
\caption[]{The $\sigma$  [X/Ca] for elements that are measured
in at least two of the nine CEMP stars (HE\,0107-5240
HE\,1327-2326, HE\,0557-4840, HE\,0134-1519, HE\,0233-0343, HE\,1310-0536,
 SMSS J0313-6708,  CS 22949-037, and G 77-61), as a 
function of atomic number. As in Fig. \ref{plotca},
C and N in  CS 22949-037 were not taken into account. 
}
\label{plotsigma}
\end{figure}
%%% FIGURE %%%%%%%%%%%%%%%%

\section{On the chemical composition and star formation\label{cc}}

We next  compare the abundances we have determined
in the four stars 
SDSS\,J0212+0137,
SDSS\,J0929+0238,
SDSS\,J1035+0641, and
SDSS\,J1742+2531  with those observed in other
CEMP stars of extremely low metallicity.
We selected from the literature the following sample 
of stars with $[{\rm Fe/H}]<-3.5$:
HE\,0107-5240 \citep{chris2004}; 
HE\,1327-2326 \citep{Frebel08}; HE\,0557-4840 \citep{norris07};
HE\,0134-1519, HE\,0233-0343, HE\,1310-0536 \citep{terese};
 SMSS J0313-6708 \citep{Keller};  CS 22949-037
\citep{depagne02,norris02}; and G 77-61   \citep{PlezCohen,PCM2005}.
In addition, as an example of a 
carbon-normal star, we consider the most
metal-poor star known to date, SDSS\,J1029+1729 \citep{Caffau12}. 
We note that  all six  stars known before this paper (see Table \ref{compfeh})  
with $[{\rm Fe/H}]<-4.5$ (ultra-iron-poor, UIP) are represented
in this sample.
The only elements that are commonly measured in all of these stars
are carbon and calcium,
but several other elements are derived in at least two of the stars.
Thus, in Fig.\ref{plotca}
we plot the [X/Ca] ratios for our program stars.
Placing all the measurements
for the remaining ten stars in the plot would make it
difficult to  read. Thus, we plot  the individual
ratios for SDSS\,J1029+1729, and for the remaining nine
CEMP stars we plot only the mean [X/Ca] value
for all elements that have been
measured in at least two of the nine stars. We represent the scatter
by drawing 2$\sigma$ bars around each mean value.
Given the differences between the different groups in the
treatment of NLTE and granulation effects, we have
plotted for all the stars the 1D LTE abundances to make them   
readily comparable and related all the abundances to the same
set of solar abundances.
For the Ca abundances we  only consider
the \ion{Ca}{ii} abundances for which departures from LTE
are negligible \citep{spiteCa,Korn09,Mash07}.
The exception is  CS 22949-037
for which the only LTE Ca abundances
available are based on \ion{Ca}{i};
we took the LTE value of \citep{depagne02}\footnote{This implies 
$[{\rm Fe/Ca}]=-0.41$; the NLTE value of
\citet{spiteCa} for the \ion{Ca}{i} subordinate lines
implies $[{\rm Fe/Ca}]=-0.69$; the NLTE \ion{Ca}{ii} of
\citet{spiteCa} implies $[{\rm Fe/Ca}]=-0.88$.}.

\subsection{Low scatter in the iron-to-calcium ratio.\label{lowca}}

We would like to note
two features apparent in Fig. \ref{plotca}: {\em i)} although the 
present measures have not been used to determine the
mean values and dispersions shown in the figure, all
the [X/Ca] values are well within the $2\sigma$ dispersion, with the
only marginal  exception of [Si/Ca] in SDSS\,J0212+0137; and 
{\em ii)} the carbon-normal star SDSS\,J1029+1729 
has not been used to determine the mean values and dispersions, 
but  its [X/Ca] values are all within $2\sigma$ of the dispersion.

\begin{figure}
\begin{center}
\resizebox{\hsize}{!}{\includegraphics[clip=true]{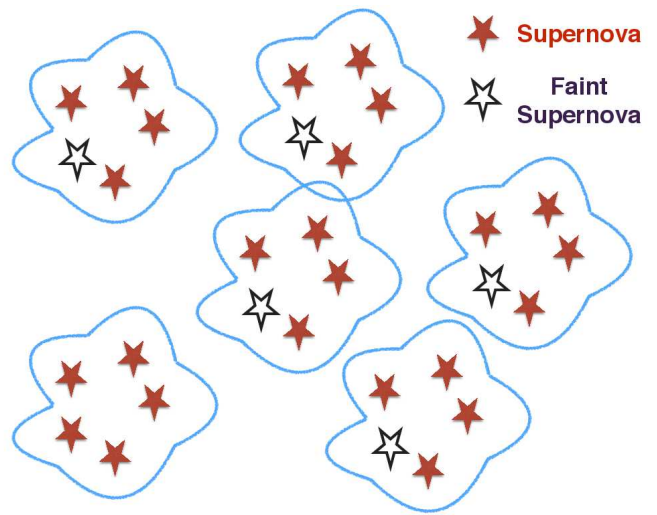}}
\end{center}
\caption[]{Scenario discussed
in Sect. \ref{lowca}. The primordial clouds produce several
massive stars, all of which explode as supernovae, but some are ``faint''
supernovae, with extensive fall-back. The supernova ejecta pollute the cloud
determining the chemical composition of the next generation of stars. 
Occasionally different star formation regions may reciprocally contaminate each
other.}
\label{SF_cartoon}
\end{figure}

In Fig.\,\ref{plotsigma} 
we show the dispersion $\sigma$ in the [X/Ca] ratios for the sample
of nine stars detailed above. 
The plot 
shows that there is a drop in the dispersion of the 
[X/Ca] ratios for all the elements heavier than Al, although the stars span   3\,dex in A(Fe). 
The scatter in the [Fe/Ca] ratio is only 0.23\,dex.
On the contrary, the lighter
elements show a much larger scatter, up to 1\,dex for carbon.
This evidence could be naturally interpreted in terms of a 
general scenario, already addressed 
by \citet{lcb2003}, and \citet{bcl-Nature},
which is based on the fact that the 
binding energy of a star at the onset of the core collapse 
increases as the metallicity decreases, the main reason being 
that these stars are more compact because of the lower opacity 
and also because of the much lower mass 
loss they experience during their lifetime. Hence, one expects  the
number of faint supernovae
to increase as the metallicity decreases. 
In this scenario, the Pop III core collapse supernovae with progenitors 
lower than $\sim 25~{\rm M_\odot}$ produced
the elemental distribution pattern shown by the normal 
extremely metal-poor (EMP; for the present discussion 
this includes all stars with $[{\rm Fe/H}]\le -3.0$) stars \citep{lc2012}. 
In the clouds already  enriched by these supernovae, 
every now and then a rather massive star exploded 
with an extended fall-back due to its large
binding energy. In this case most of the heavy elements produced in the innermost layers
remained locked in the compact remnant and only the 
lighter elements were mixed in the surrounding. 
These ejecta were responsible for  the variety of 
abundance patterns shown by the light elements in most of the presently known 
UIP stars ($[{\rm Fe/H}]<4.5$).
The scenario is depicted  in Fig. \ref{SF_cartoon}.

%%% FIGURE %%%%%%%%%%%%%%%%
%%% FIGURE %%%%%%%%%%%%%%%%

It is worth noting that the abundance pattern produced by
explosive burning does not show a strong dependence 
on the initial mass of the progenitor,
so that even very different initial mass functions (IMFs) would produce a similar 
abundance pattern: at the extreme, even the ejecta of a single 
supernova could have been responsible of the
observed pattern above Mg \citep{LC02}. 
No stars at metallicity below --4.5 have yet been detected outside
the Galaxy; however, there are now a number of very metal-poor stars
(i.e. below --2.5)
that are known in Local Group Galaxies, and many are below --3.0. 
To date, the most metal-poor star known in external
galaxies is found in Sculptor and has a metallicity near --4.0
\citep{TAF10};  two more stars of similar metallicity
are  known in  Sculptor \citep{anna10,Simon15} and six more
below --3.0 \citep{Starkenburg13,Simon15}. 
Other local galaxies with at least one such star
are Draco \citep{Shetrone2001,Fulbright2004,CH09}, Sagittarius
\citep[][and Monaco et al. in preparation]{Sgr04,Sgr06,Sbordone15},
Fornax \citep{TAF10}, Ursa Minor \citep{KC12,ural},
Sextans \citep{Aoki09,TAF10}, Coma Ber \citep{anna10b}, Ursa Maior II 
\citep{anna10b}, Segue 1 \citep{RK14,anna14}, Hercules \citep{K08,Aden},
Boo I \citep{Gil13,Ishi14}, and Boo II \citep{KR14}.
Particularly relevant to the present discussion is the discovery
of a CEMP-no star in the Sculptor dwarf spheroidal \citep{SK15}.
The conclusion of the authors is that the fraction of CEMP-no stars
in Sculptor seems to be significantly lower than in the Galactic Halo.
In our scenario this could be explained either with a lower frequency
of faint SNe in Sculptor, or with shallower potential wells
that can allow the formation of the second generation of stars.
Another CEMP-no star of extremely low metallicity
was previously discovered in Bootes I 
\citep{Lai2011,Gil13} and  this one also lies in 
the {\em low-carbon band}. 
In any case the presence of at least  two CEMP-no stars outside our Galaxy
prompts us to search for a formation mechanism  that is universal.
Clearly, further scrutiny of the EMP populations in Local Group
galaxies, as can be afforded by multi-object spectrographs on 
30\,m class telescopes \citep{mosaic}, is of the highest interest.

We  also note that
the   upper limit for the [Fe/Ca] ratio in SMSS J0313-6708  
is consistent with the
values observed in the other stars. If the iron abundance in this star
were measured to be up to 0.4\,dex below the current upper limit 
it would be similar to the others; only if it were much lower  would it be definitely peculiar.
 If our scenario is correct the ratios of
all elements heavier than Si to Ca in SMSS J0313-6708
ought to be within the mean trends shown in Figs. \ref{plotca}
and \ref{plotsigma}, although  its [C/Ca] and [Mg/Ca] ratios
are at several $\sigma$ from the mean.

\subsection{Lithium}
%%% FIGURE %%%%%%%%%%%%%%%%
\begin{figure}
\begin{center}
\resizebox{\hsize}{!}{\includegraphics[clip=true]{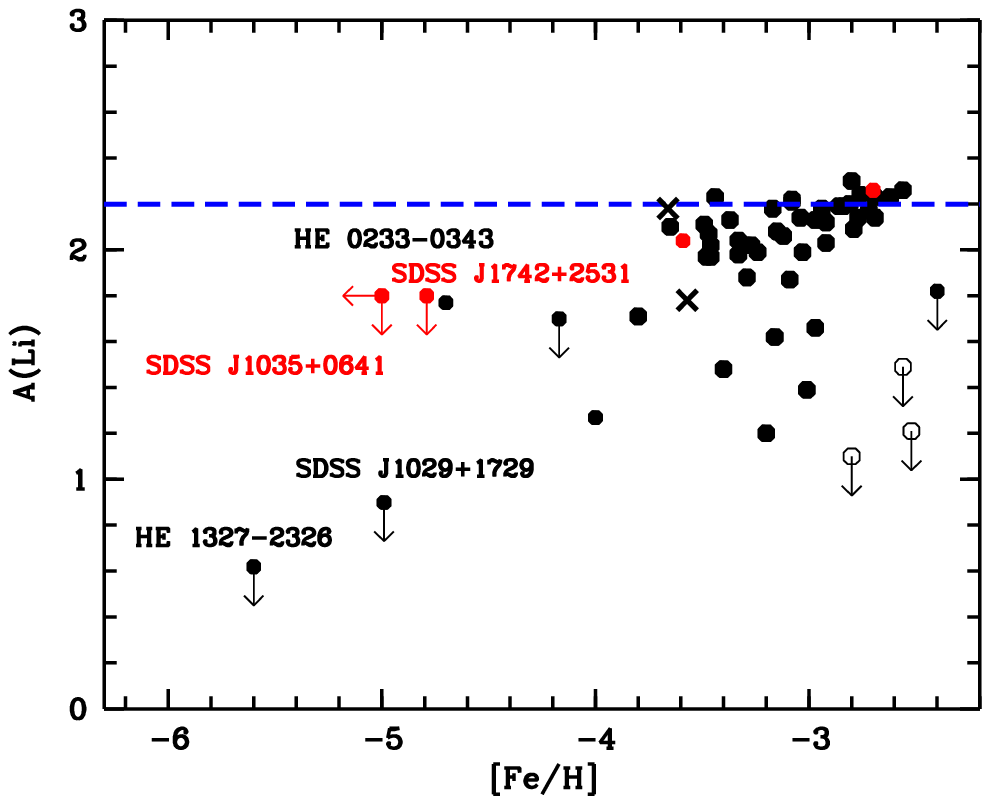}}
\end{center}
\caption[]{
Lithium abundances, or significant upper limits, 
in the stars analysed in the present paper (red symbols),
compared to those in other extremely metal-poor
unevolved stars \citep{Frebel08,sbordone10,bonifacio12,Caffau12,terese} (black symbols).
The two components of the binary system CS 22876-32 \citep{jonay08} are shown as crosses.
The upper limits shown as open symbols are G 186-26, G 122-69, and
G 139-8 from \citet{norris97a}. 
The blue dashed line is the level of the {\em} Spite plateau
as determined by \citet{sbordone10}.
}
\label{plotli}
\end{figure}
%%% FIGURE %%%%%%%%%%%%%%%%

There  are now five known unevolved stars with
iron abundance below $-4.5$:  HE\,1327-2326 \citep{frebel05},
SDSS\,J1029+1729 \citep{Caffau12}, 
HE\,0233-0343 \citep{terese}, SDSS\,J1742+2531 (\citealt{gto13} and this paper)
and SDSS\,J1035+0641 (this paper).
SDSS\,J1742+2531 and SDSS\,J1035+0641  have no measurable lithium, 
like HE\,1327-2326 and SDSS\,J1029+1729,
at variance with most
of the stars found at higher iron abundance with similar
atmospheric parameters.  
The situation is shown in Fig. \ref{plotli}.
HE\,0233-0343 is very similar
to SDSS\,J1742+2531 both for atmospheric parameters and
chemical abundances of C, Ca, and Fe, but it has a detectable
lithium doublet that implies A(Li)=1.77 \citep{terese}, thus roughly
a factor of three below the Spite plateau. 

The Spite plateau meltdown, highlighted by
\citet{aoki09} and \citet{sbordone10}, is the large number of stars
at metallicity below $-3.0$ solar, which have lithium
abundances lower than the Spite plateau, but still measurable
\citep[see figure 3 in][]{spitemem}.
In  three of the above-mentioned UIP stars
the lithium  abundance could not be measured, and it is
intriguing to consider the possibility that the atmospheres
of these stars are indeed totally devoid of lithium.
How did this happen? Is this related to the very low
iron abundance of these stars?
We note that owing to the difference in the C abundances, their
total metallicity $Z$ is indeed very different.

The observation of HE\,0233-0343 implies that the phenomenon
responsible for the total destruction of lithium is not
effective in all cases. It is, however, natural to suppose that it
is  the same  phenomenon that is responsible for the
lithium depletion in HE\,0233-0343 and for the 
Spite plateau meltdown.

A possible explanation  is that
all these stars were  born with high rotational
velocities as a consequence of the 
role of fragmentation in their formation, as argued 
in section \ref{sf} and as supported
by the simulations of \citet{stacy11,stacy13}. 
We note that the gas out of which the first stars form was likely to be strongly magnetised. 
Any pre-existing magnetic field was amplified to dynamically significant levels by the 
small-scale turbulent dynamo which converts parts of the kinetic energy associated with 
the accretion flow in the star-forming parts of the halo into magnetic energy 
\citep{schleicher10,sur10,schober12}.
In principle, magnetic fields can remove angular momentum from the system by magnetic braking 
\citep[e.g.][]{mouschovias85,hennebelle09}.
However, the efficiency of this process is strongly reduced in turbulent fields with a highly tangled morphology 
\citep{seifried12,seifried13},
and so an appreciable fraction of the angular momentum is thought to remain in the system. 
This --together with the fact that primordial stars are expected to have very weak winds 
\citep[][]{ekstrom08}-- 
supports the hypothesis that these stars remain fast rotators at least during all 
of their pre-MS and a good part of their MS phase 
\citep{maeder12}. Rotational mixing would then bring the material below
the region of Li-burning, leading to the total destruction of 
Li in these stars.  
As the metal content increases the star formation mode shifts
to a more standard Pop II formation mode, with lower
rotational velocities and less Li destruction.
In this scenario the 
Spite plateau meltdown would simply mark the transition
between the Pop III and Pop II star formation mode.

A second, related,   scenario has been proposed
by \citet{molaro}. They assume  Li is  significantly 
destroyed in the pre-MS phase by overshoot mixing, 
and then partially restored by late accretion of
fresh non-Li depleted material. The extreme objects, 
such as 
SDSS\,J1742+2531, SDSS\,J1029+1729, SDSS\,J1035+0641 
and HE\,1327-2326, simply lacked the late accretion phase 
because they formed by fragmentation,   and as a result 
the proto-star was ejected from the gas reservoir.

Finally, we wish  to propose a third possible
scenario that does not invoke high rotation in 
the early stages of the star's life.
The first stars formed in dark-matter  minihaloes
\citep[see][and references therein]{Bromm13}.
The concept of minihaloes that eventually merge
to form a galaxy is at the heart of the cold dark matter, with Cosmological constant $(\Lambda$-CDM)
  scenario
of hierarchical galaxy formation.
If all the first-generation stars explode as
SNe, with typical energies  of $10^{51}$ ergs,
then the remaining gas in the minihalo is blown out
and no further star formation can take place.
However, in some minihaloes a few typical  SNe explode
in addition to  faint SNe. In this case the gas is not
blown out and a second generation of stars may form.
The second-generation stars may have
both a CEMP-no chemical composition or a normal
EMP composition, depending on the mass of the  ejecta
contributed by  faint and normal SNe. 
However, in all these stars the Li content
would be lower than that present in the
primordial gas. The SNe ejecta being Li-free, the resulting
Li abundance would simply depend on the amount of dilution
of the SN ejecta with pristine material.
If we assume that many of the stars with [Fe/H]$\la -2.8$ have been 
formed in such minihaloes, this could explain the Spite plateau
meltdown. In this case the most metal-poor stars would simply
be an extreme case of stars formed from ejecta very little diluted
with pristine material.
Once the galaxy is assembled, through the merging of many minihaloes,
a standard chemical evolution, with well-mixed matter begins
\citep{Prantzos}. The final result is that all the Li-free material 
of the SNe ejecta dilutes the primordial lithium abundance. 
This could solve the cosmological Li problem.

In all three  of the above scenarios
the case of HE\,0233-0343 could be accommodated if we 
assume that the phenomenon has various degrees of efficiency.
In the case of rotationally induced mixing, this should 
be partially inhibited by some kind of braking (magnetic or other).
In the case of late
accretion, this should be allowed to proceed long enough to 
restore the lithium to the observed abundance.
In the case of mixing of primordial matter with SNe
ejecta in minihaloes, it is the mass ratio of the two
materials that governs the final Li abundance.

\subsection{Implications on the formation of low-mass stars in metal-poor gas\label{sf}}

Star formation in metal-free or extremely metal-poor environments is more complex than previously thought. 
While early simulations of the formation of primordial stars suggested
masses above 100$\;$M$_{\odot}$ 
\citep{OP2001,OP2003,Abel2002,Bromm2002},
more recent work in early star formation acknowledges the importance of
turbulence 
\citep{clark11,greif11a,greif11},
variations of halo properties \citep{Hosokawa,Hirano}, 
cosmological streaming velocities \citep{TH10,greif11a,Maio11,stacy11b,TH11},
and even magnetic fields  \citep{Schleicher,Peters}
or potentially dark-matter annihilation \citep{Iocco08,Spolyar09,Smith12}.
All these processes influence  the fragmentation properties of the star
forming gas and consequently modify the resulting stellar mass spectrum. As discussed in Section  \ref{chigh},
current studies predict a wide range of masses all the way down into the
substellar regime. However, a full consensus on the IMF of primordial as well 
as of extremely metal-poor stars has yet to emerge, and so further investigations are required. As
soon as the metallicity rises above a certain threshold, which is thought to lie between $10^{-6}$ and $10^{-5}\;Z_{\odot}$ 
\citep[e.g.][]{omukai05,schneider12,Glover13},
two main cooling channels emerge which could lead to the efficient formation of low-mass stars and result 
in a stellar mass spectrum similar to the present-day IMF
\citep{Kroupa02,Chabrier03}. These cooling processes are

\begin{itemize}
\item
fine structure transitions of \ion{C}{ii}
and \ion{O}{i} \citep{brom_loeb} and
\item
dust cooling 
\citep{schneider06,omukai08,dopcke11,schneider12,dopcke13}.
\end{itemize}

 SDSS\,J1742+2531 is the seventh star discovered
with $[{\rm Fe/H}]<-4.5$ and out of seven, six are CEMP stars (on the {\em low-carbon band})
and only one,
SDSS\,J1029+1729, is a carbon-normal star.

The {\em low-carbon band} places all these CEMP stars
in the ``permitted zone'' of the
\citet{brom_loeb} theory, thus making formation via
metal-line cooling viable. 
This is not the case 
for SDSS\,J1029+1729,
for which dust cooling appears necessary
\citep{schneider12b,klessen12,chiaki}.

It would be tempting to conclude, as  \citet{Norris13} and
\citet{Gil13} did, that
the CEMP stars of the {\em low-carbon band} 
were formed via line cooling, while
SDSS\,J1029+1729   formed via dust cooling.
However, even for the C-enhanced gas, 
line cooling only lowers the
Jeans masses (and as a consequence
stellar masses) to 
values largely in excess of 10 M\sun. 
Clearly, in order to form the low-mass
objects observed, 
one would need to resort to a high-density cooling process, such as 
dust cooling. 
This is the only process that brings 
the gas into the right regime of masses  consistent
with the observations \citep{omukai05,clark08,klessen12,schneider12,schneider12b,dopcke13,chiaki,chiaki15}.
It seems thus more realistic to assume that all the stars
in Fig.\,\ref{plotca} were formed by dust cooling,
irrespective of their carbon content.

We note, however, that
the above arguments are based solely  on thermodynamics. 
If we  also consider  the  dynamics of the collapsing cloud
it turns out that even completely metal-free gas can 
make low-mass objects simply by disc fragmentation 
\citep[e.g.][]{clark11,greif11,greif12} because  the gas is almost
isothermal over a wide range of 
densities\footnote{The equation of state can be expressed as 
$P \propto \rho^\gamma$ with  $ 1.05\la \gamma \la 1.1$ and $\rho$ density, with no explicit dependence
on temperature.}.
The mass load onto the accretion disc exceeds its ability to transport material
inwards, and it fragments. This fragmentation is not as widespread as with dust cooling.
The expected IMF is flat \citep{dopcke13}, which means that it 
is still  top-heavy  with respect
to  today's standards, but 
it is possible to  make low-mass objects   in the mass range of the stars in 
Fig.\,\ref{plotca}. Therefore,  in principle, 
all these stars may have been formed without resorting to either dust or metal-line cooling.

In a recent paper, \citet{cooke14} proposed a numerical 
model of the chemical enrichment produced by the first
generation of stars, and looked at the resulting chemical composition  
irrespective of the mechanism
needed to form the next generation of low-mass
stars. Their model has several  points in common
with our proposed scenario. In particular, they assume
that the first stellar generations are formed in 
minihaloes of a mass of a few millions of solar
masses. In these haloes one or several massive stars are formed. 
\citet{cooke14} recognise the important link between the
explosion energy of the resulting supernovae and the ability
of the minihalo to retain gas and thus form a new
generation of stars. The consequence is that the minihaloes that
host faint (low energy) SNe are  those that are more likely
to form second-generation stars, and because of the presence
of the faint SNe these are indeed C-enhanced.
This model reproduces well the increase in the fraction 
of CEMP stars at low metallicity and the role of
faint SNe is very similar to the scenario that we propose.
However, the model is less successful in reproducing the
detailed abundances of the EMP stars. 
In particular at the lowest metallicities
all the stars seem to lie at [C/Fe] values much
higher than those preferred by the models. 
Here  we have presented data for four more such stars, 
suggesting that these high [C/Fe] values, which correspond 
simply to the {\em low-carbon band} discussed here, are the norm
at low metallicities and not the exception. 
This rules out exotic explanations like enrichment
from AGB stars or CNO boosted nucleosynthesis 
\citep[see][and references therein]{cooke14}.
We believe that the  most likely possibility 
is the one proposed by \citet{cooke14}, that the
mixing of the SNe ejecta is not uniform.
In Sect. \ref{lcband} we suggest that the existence of the
{\em low-carbon band} provides a strong
constraint to the minimum dilution of SNe ejecta necessary for
star formation. 
In the context of the \citet{cooke14} models we argue that the {\em \emph{width}} 
of the band should provide a constraint on the dispersion in dilutions
encountered.  
It would be interesting to run models of the type of
the  \citet{cooke14} models, with a variable dilution to see
if it is possible to reproduce the {\em low-carbon band}.

\subsection{Alternative scenarios}

There are, of course, alternative scenarios to our preferred 
faint supernovae scenario that may explain the occurrence
of CEMP stars of extremely low metallicity.
We refer the reader to the introduction of the recent
paper by \citet{MMC} who list as many as five possible
scenarios  and to the extensive
review of \citet{Karlsson} on the chemical signatures of the first stars.
Here we wish  only to mention the one that we believe
is the most serious competitor to the faint supernova
scenario: rapidly rotating massive stars.
A rapidly rotating massive star may,
during its main sequence lifetime, bring  
C and N to the surface. 
\citet{PCM2005} invoked this mechanism to explain 
the large N abundance and low $^{12}$C/$^{13}$C ratio
in one of the first CEMP stars of extremely low metallicity 
discovered: G\,77-61.
The role of rapidly rotating
massive stars in producing CEMP stars 
has often been discussed 
in the literature 
\citep{Meynet06,Chiappini06,Hirschi07,Meynet10,MMC}.
In the presence of rapid rotation, N in particular would be
greatly increased by rotation during the central He burning phase when a
tail of C produced by the He burning is transported to the base of the H
burning shell by rotationally induced mixing, and then rapidly converted into N by
the CNO cycle. In particular, we expect that a generation of stars rotating
initially at 300 km/s, integrated over a Salpeter IMF, would produce ejecta
(but not necessarily wind) with a C/N ratio of the order of 17 (Limongi and
Chieffi in preparation), while the analogous generation of non-rotating stars
would provide a C/N ratio of the order of 3500. The same holds for the
$^{12}$C/$^{13}$C  ratio that turns from 350000 to 2000 from v=0 to v=300. It is worth
noting that rotation should not be extreme, in the sense that it must not
lead to a full mixing of the star because some active H burning must be
still present when C is produced by the central He burning.

In the future we plan to make a detailed comparison of the predictions
of faint supernovae and rapidly rotating massive stars.

\section{Conclusions}

We have determined the iron abundance
in SDSS\,J1742+2531 to be $[{\rm Fe/H}]=-4.8$. 
This increases
 the sample of known stars with $[{\rm Fe/H}]<-4.5$ to seven. 
We could not measure the iron abundance in SDSS\,J1035+0641, yet
our  3$\sigma$ upper limit of $-5.00$ classifies it as the eighth
such star, supported also by its extremely low Ca abundance ($[{\rm Ca/H}]=-5.0$).
Both stars are carbon enhanced (CEMP), thus the sample of the eight most
iron-poor stars is composed of seven CEMP stars and one 
carbon-normal star (SDSS\,J1029+1729).

Based on the measured [X/Ca] ratios in a sample of  CEMP stars, 
and theoretical predictions of 
nucleosynthesis in zero-metallicity supernovae, we have proposed
a scenario in which the metals observed in these stars
were produced by  a few zero-metallicity SNe.  
The carbon abundances of the program stars confirm the bimodal
distribution in CEMP stars, with the CEMP-no stars occupying
a {\em low-carbon band} identified by \citet{Spite13}. We have proposed
an interpretation of this bimodality: the stars on the {\em high-carbon band},
all of which are binaries \citep{Starkenburg},
are the  result of mass-transfer from an AGB star in a binary
system, while the {\em low-carbon band} stars 
reflect the abundances of the gas out of which they were formed,
the high-carbon abundance is the result of the ejecta
of a faint supernova.

This scenario
is also supported by the theoretical predictions
that the zero-metallicity stars are not formed in isolation
but are formed in groups, and possibly as 
multiple stars \citep{clark08,Stacy10,clark11,clark11b,greif11,greif12,SB13}. 
The idea that several massive Pop III stars may pollute a primordial
cloud, and that subsequently
low-mass stars can form in the high-density
shocks was already proposed by \citet{cayrel86}.
However, at that time no stars as metal-poor as those discussed here
were known, and it was suggested
that this mechanism could enrich the gas up
to one hundredth of the solar value or even above.
Our proposed scenario, following the ideas
of \citet{bcl-Nature} and \citet{lcb2003}, requires a few
SNe of mass less than $\sim 25$M$_\sun$.
Since our scenario requires more than one SNe, it
departs significantly from other attempts
to explain the abundance pattern of these stars by 
the ejecta of a single exotic SN 
\citep[see e.g.][and references therein]{umeda,iwamoto,izutani, joggerst,ishigaki}.
Our scenario explains
well the low star-to-star scatter in the [X/Ca] ratios  (a factor of two, at most) for elements
heavier than Mg, in spite of the large span in [Fe/H] displayed
by these  stars.
The large scatter in the [X/Ca] ratios for the lighter
elements, and in particular the CEMP nature of some of the stars,
are determined by the presence, among the SNe that produced the metals,
of at least one faint SN. The fact that seven out of eight stars
found to have [Fe/H] below --4.5 are CEMP, suggests that the occurrence of 
the faint SNe is a rather common event  probably because  the binding energy of the
lower metallicity stars at the onset of
the core collapse is larger, thus favouring  fall-back,
and  the faintness of the SN.
The statistics are clearly scanty and before drawing conclusions
on the frequency of faint SNe one should study possible selection
biases. This is not trivial given that the seven most iron-poor
stars have been selected from three different surveys (Hamburg ESO, SDSS,
and SMSS) with different follow-up procedures. The fact that three
of the eight stars have been selected by our group from SDSS
using the same criteria and include one carbon-normal  (SDSS\,J1029+1729)  
and two CEMP (SDSS\,J1035+0641 and SDSS\,J1742+2531) suggests that there is no
strong bias in favour of or against the selection of CEMP stars.

Very interestingly, SDSS\,J1742+2531 and SDSS\,J1035+0641  do not show any detectable
lithium, and this characteristic is shared by two of the other 
unevolved stars of the sample (SDSS\,J1029+1729 
and HE\,1327-2326). We proposed three possible scenarios
to explain these Li-poor stars.
In two of the scenarios 
they were formed by fragmentation and this  resulted
in the destruction of lithium in the pre-main-sequence phase, either
through rotational mixing or owing to the lack of the late
accretion phase as proposed by \citet{molaro}. 
A third scenario requires that these stars were formed within
a minihalo, in which the Li-free SN ejecta were mixed with 
primordial gas. Their formation must have taken place before
the minihaloes merged to form the Galaxy.
In all three cases
these stars 
 had a different formation with respect to the stars
with higher [Fe/H]. The case of HE\,0233-0343, which is very
similar  to SDSS\,J1742+2531 except for the lithium abundance, 
implies that even in this star formation mode some lithium can
survive. It is, however, remarkable that the lithium observed
in this star is still almost a factor of three below the Spite
plateau, and similar to the value observed in many of the stars
of the Spite plateau meltdown.

Among the different mechanisms for the formation of these
UIP stars, the \citet{brom_loeb} cooling mechanism is viable
for all the CEMP stars, yet it still requires  either
a fragmentation
of the gas cloud at the late stages of the collapse or an additional 
high-density cooling mechanism.
On the other hand, all forms of dust cooling 
\citep{schneider06,omukai08,dopcke11,schneider12,dopcke13}
are viable for all the UIP stars, CEMP or not.
Disc fragmentation 
\citep[][]{clark11,greif11,greif12}, is also
a possible formation mechanism for all these stars, and it
would also allow the formation of primordial low-mass stars.

The indirect evidence coming from the lack of lithium in three of
the four  unevolved
stars,  suggests a prominent role of disc fragmentation, either in the
presence of dust or not.
A direct confirmation of this mechanism would be the detection of a low-mass
star totally devoid of metals. The search of such stars must continue,
and an interesting prospect is offered by an all-sky, unbiased survey
such as Gaia\footnote{\url{http://sci.esa.int/gaia/}}.

%%%%%%%%%%%%%%%%%%%%%%%%%%%%%%%%%%%%%%%%%%%%%%%%%%%%%%%%%%%%%%%%%%%

\begin{acknowledgements}
The project was funded by FONDATION MERAC.
PB, EC, PF, MS, FS, and RC acknowledge support from the Programme National
de Cosmologie et Galaxies (PNCG) and Programme  National de Physique Stellaire
of the Institut National de Sciences
de l'Univers of CNRS.
EC, RKS, HGL, NC, SCOG, and LS acknowledge financial support
by the Sonderforschungsbereich SFB881 ``The Milky Way
System'' (subprojects A4, A5, B1, B2 and B8) of the German Research Foundation
(DFG).
ML and AC acknowledge financial support from PRIN MIUR 2010-2011, project ``The
Chemical and dynamical Evolution of the Milky Way and Local Group
Galaxies'', prot. 2010LY5N2T.
RSK furthermore acknowledges support from the European Research Council under the 
European Community’s Seventh Framework Programme (FP7/2007-2013) via the ERC Advanced Grant STARLIGHT (project number 339177).
LS acknowledges the support of  Project IC120009 "Millennium Institute of Astrophysics (MAS)" of Iniciativa Cient\'ifica Milenio del Ministerio de
Econom\'ia, Fomento y Turismo de Chile
EC and PM acknowledge support from
the international team \#272 lead by C. M. 
Coppola EUROPA- Early Universe: Research On Plasma Astrochemistry" at ISSI (International Space Science
Institute) in Bern.

\end{acknowledgements}

\bibliographystyle{aa}

\Online
\begin{appendix}
\section{Ratios of various elements to calcium}

For the reader's convenience we present in the following table
the data that were used to plot Fig. \ref{plotca}.
All the published ratios have been scaled to bring them to
our adopted solar abundance scale.
As solar reference for carbon, nitrogen, oxygen, and iron we took \citet{abbosun},
and for the other elements \citet{lodders09}.

\begin{table*}
\caption{\relax [X/Ca] ratios for the four  low-carbon band stars studied in the present paper 
and for the carbon-normal most metal-poor star SDSS J1029+1729.}\label{stars}
 \centering
\begin{tabular}{rrccccc} 
\hline\hline 
% Table : stars 
%Sequence    A                 Ceti Hydra      Leo2   Ercoli    Caffau 
% --------    ----  ------ ------ -------------- ----
                        & Z &   (1)  &  (2)   &  (3) &  (4)  &  (5)\\           
\hline                                                
\relax        [C/Ca]    & 6 &  +1.43 & +3.23  &+3.38&  +3.32&     -  \\
\relax        [N/Ca]    & 7 &  +1.85 &     -  &    -&      -&     -  \\
\relax        [O/Ca]    & 8 &  +0.75 &     -  &    -&      -&     -  \\
\relax        [Na/Ca]   &11 &      - &     -  &    -&      -&     -  \\
\relax        [Mg/Ca]   &12 &  -0.35 &     -  &    -&      -& -0.11  \\
\relax        [Al/Ca]   &13 &  -1.38 &     -  &    -&      -&     -  \\
\relax        [Si/Ca]   &14 &  -0.39 &     -  &    -&      -& +0.33  \\
\relax        [Sc/Ca]   &21 &  -0.29 &     -  &    -&      -&     -  \\
\relax        [Ti/Ca]   &22 &  -0.10 &     -  &    -&      -& -0.15  \\
\relax        [Cr/Ca]   &24 &  -0.74 &     -  &    -&      -&     -  \\
\relax        [Mn/Ca]   &25 &      - &     -  &    -&      -&     -  \\
\relax        [Fe/Ca]   &26 &  -0.78 &     -  &    -&  -0.24& -0.13  \\
\relax        [Co/Ca]   &27 &  +0.38 &     -  &    -&      -&     -  \\
\relax        [Ni/Ca]   &28 &  -0.65 &     -  &    -&      -& +0.05  \\
\relax        [Sr/Ca]   &38 &  -1.02 &     -  &    -&      -&     -  \\
\relax        [Ba/Ca]   &56 &  -0.62 &     -  &    -&      -&     -  \\
% --------    ---- ------- ---- ------ --- ------ ----- ------------  ------ 
\hline
\multicolumn{7}{l}{(1) SDSS J0212+0137  This paper}\\
\multicolumn{7}{l}{(2) SDSS J0929+0238  This paper}\\
\multicolumn{7}{l}{(3) SDSS J1035+0641  This paper}\\
\multicolumn{7}{l}{(4) SDSS J1742+2531  This paper}\\
\multicolumn{7}{l}{(5)  SDSS J1029+1729 \citet{Caffau12}}\\
\hline
\end{tabular}
\end{table*}

\begin{sidewaystable*}
\caption{The
\relax [X/Ca] ratios  
for all the known CEMP stars with \relax [Fe/H] less than about --4.0.
All the published
abundances have been scaled to our adopted solar values.}\label{stars1}
 \centering
\begin{tabular}{rrccccccccccccc} 
\hline\hline 
% Table : stars 
                        & Z &   (1)  &  (2)  &  (3) &  (4)  &  (5)  &  (6)  &  (7)   & (8)  &  (9)    & Mean & $\sigma$ & Min. & Max. \\           
\hline                                        
\relax        [C/Ca]$^a$    & 6 & +3.20 & +0.82 & +3.06 & +4.72 &  +2.53 &+0.94  &  +2.28 &+3.39 &  +0.82 &+2.44 & +1.04 & +0.82 & +3.39  \\
\relax        [N/Ca]$^a$    & 7 & +2.25 & +0.86 &     - &     - &  +1.57 &    -  &  +3.16 &+3.92 &  +2.26 &+2.35 & +1.22 & +0.86 & +3.92  \\
\relax        [O/Ca]    & 8 & +1.79 &     - &     - &     - &  +1.57 &    -  &      - &+3.07 &  +1.68 &+2.18 & +0.77 & +1.68 & +3.07  \\
\relax        [Na/Ca]   &11 & +0.45 & -0.41 &     - &     - &  +0.03 &-0.89  &  +0.12 &+1.68 &  +1.10 &+0.33 & +0.87 & -0.89 & +1.68  \\
\relax        [Mg/Ca]   &12 & -0.24 & +0.20 & +0.30 & +2.95 &  -0.11 &-0.36  &  +0.47 &+0.99 &  +1.23 &+0.26 & +0.53 & -0.36 & +1.23  \\
\relax        [Al/Ca]   &13 &     - &     - &     - &     - &      - &-1.31  &  -0.42 &+0.42 &  -0.53 &-0.56 & +0.73 & -1.31 & +0.42  \\
\relax        [Si/Ca]   &14 &     - & -0.07 & +0.01 &     - &      - &    -  &      - &    - &  +0.13 &+0.10 & +0.17 & -0.07 & +0.33  \\
\relax        [Sc/Ca]   &21 &     - & -0.16 &     - &     - &      - &-0.34  &  -0.18 &    - &  -0.20 &-0.16 & +0.17 & -0.34 & +0.07  \\
\relax        [Ti/Ca]   &22 & -0.63 & +0.05 & -0.12 &     - &      - &-0.25  &  +0.39 &-0.00 &  +0.11 &-0.10 & +0.26 & -0.63 & +0.18  \\
\relax        [Cr/Ca]   &24 &     - & -0.33 &     - &     - &  -0.21 &-1.30  &  -0.50 &    - &  -0.75 &-0.53 & +0.51 & -1.30 & -0.01  \\
\relax        [Mn/Ca]   &25 &     - & -1.24 &     - &     - &  -0.64 &    -  &  -1.35 &    - &  -1.16 &-1.05 & +0.41 & -1.35 & -0.44  \\
\relax        [Fe/Ca]   &26 & -0.57 & -0.13 & -0.37 &     - &  -0.67 &-0.60  &  -0.03 &-0.74 &  -0.41 &-0.39 & +0.21 & -0.74 & -0.13  \\
\relax        [Co/Ca]   &27 &     - & +0.21 &     - &     - &      - &-0.57  &  +0.16 &    - &  -0.04 &-0.06 & +0.36 & -0.57 & +0.21  \\
\relax        [Ni/Ca]   &28 & -0.74 & +0.07 &     - &     - &      - &-0.82  &  -0.14 &-0.51 &  -0.44 &-0.34 & +0.38 & -0.82 & +0.07  \\
\relax        [Sr/Ca]   &38 &     - & -0.46 & -0.08 &     - &      - &    -  &  -1.14 &+0.31 &  -0.04 &-0.28 & +0.55 & -1.14 & +0.31  \\
\relax        [Ba/Ca]   &56 &     - &     - &     - &     - &      - &    -  &  -0.50 &    - &  -0.99 &-0.75 & +0.35 & -0.99 & -0.50  \\
% --------    ---- ------- -------- ------ ------------- ------ ------- ----- -------  -----  ------ ------ ------ ------ 
\hline
\multicolumn{15}{l}{(1) HE 0107-5240    \citet{chris2004}}\\
\multicolumn{15}{l}{(2) HE 0134-1519    \citet{terese}}\\
\multicolumn{15}{l}{(3) HE 0233-0343      \citet{terese}}\\
\multicolumn{15}{l}{(4) SMSS J0313-6708   \citet{Keller}}\\
\multicolumn{15}{l}{(5) G 77-61            \citet{PlezCohen,PCM2005}}\\
\multicolumn{15}{l}{(6) HE 0557-4840    \citet{norris07}}\\
\multicolumn{15}{l}{(7) HE 1310-0536    \citet{terese}}\\
\multicolumn{15}{l}{(8) HE 1327-2326     \citet{Frebel08}}\\
\multicolumn{15}{l}{(9) CS 22949-037     \citet{depagne02,norris02}}\\
\multicolumn{15}{l}{$^a$ For these elements the abundances of  CS 22949-037 have not been considered to compute the mean and $\sigma$}\\
\multicolumn{15}{l}{\phantom{$^a$} since the star is ``mixed'', in the sense of Spite et al. (2005). }\\
\hline
\end{tabular}
\end{sidewaystable*}

\begin{table*}
\caption{\relax Abundances of Fe, Ca, and C for the sample
of nine comparison stars. 
}\label{compfeh}
 \centering
\begin{tabular}{rrrr} 
\hline\hline 
Star & [Fe/H] & [Ca/H] & A(C) \\
\hline\hline
HE 0107-5240    &$-5.46 $&$-4.89$ & 6.81 \\
HE 0134-1519    &$-4.00 $&$-3.87$ & 5.45 \\
HE 0233-0343    &$-4.70 $&$-4.33$ & 7.23 \\
SMSS J0313-6708 &$<-7.32$&$-7.19$ & 6.03 \\
G 77-61         &$ -4.10$&$-3.43$ & 7.60 \\
HE 0557-4840    &$ -4.75$&$-4.15$ & 5.29 \\
HE 1310-0536    &$ -4.17$&$-4.14$ & 6.64 \\
HE 1327-2326    &$ -5.73$&$-4.99$ & 6.90 \\  
CS 22949-037    &$ -4.01$&$-3.60$ & 5.72 \\
\hline
\end{tabular}
\end{table*}
\end{appendix}

\end{document}